\documentclass[aps,prd,reprint,showpacs,showkeys,nofootinbib,floatfix]{revtex4-1}
\usepackage{amsmath}
\usepackage{amssymb}
\usepackage{bm}		
\usepackage{colortbl}	
\usepackage{dcolumn}	
\usepackage{graphicx}

\newcommand{\SubroutineLibrary}[1]{\textsc{#1}}
\newcommand{\subroutine}[1]{\texttt{#1}}

\newcommand{\etal}{{\it et~al.\hbox{}\/}}

\newcommand{\Cplusplus}{\hbox{C\raise.2ex\hbox{\footnotesize ++}}}
\renewcommand{\P}[1]{\phantom{#1}}
\newcommand{\Z}{\phantom{0}}

\newcommand{\thalf}{\tfrac{1}{2}}

\newcommand{\actual}{\text{(actual)}}

\newcommand{\internal}{\text{(internal)}}
\newcommand{\DMW}{\text{DMW}}

\newcommand{\fmr}{\text{fmr}}
\newcommand{\reg}{\text{reg}}
\newcommand{\rp}{\text{(r-p)}}
\newcommand{\num}{\text{num}}

\newcommand{\rms}{\text{rms}}
\newcommand{\Schw}{\text{Schw}}
\newcommand{\self}{\text{self}}

\newcommand{\tail}{\text{tail}}
\newcommand{\thresh}{\text{thresh}}
\newcommand{\trial}{\text{(trial)}}

\newcommand{\C}{\mathsf{C}}
\newcommand{\E}{\mathcal{E}}
\renewcommand{\L}{\mathcal{L}}
\renewcommand{\O}{\mathcal{O}}
\newcommand{\Q}{\mathcal{Q}}

\newcommand{\boxop}{\Box}
\newcommand{\del}{\nabla}
\newcommand{\eqdef}{\equiv}

\newcommand{\Realpart}{\mathop{\text{Re}}}

\newcommand{\normalize}[1]{\overline{#1}}	
\newcommand{\sun}{\odot}
\newcommand{\ltsim}{\lesssim}
\newcommand{\gtsim}{\gtrsim}
\newcommand{\llbrack}{\lbrack\!\lbrack}
\newcommand{\rrbrack}{\rbrack\!\rbrack}

\newcommand{\ok}{$\surd$}			
\newcommand{\okrp}{\underline{$\surd$}}	

\newcommand{\SciNum}[2]{#1 \,{\times}\, 10^{#2}}

\newcommand{\CenterWithSizeOf}[2]{{\setbox0=\hbox{#2}\hbox to\wd0{\hss{#1}\hss}}}

\newcommand{\CenterInSciNumSpace}[3]{\phantom{#2} \, \CenterWithSizeOf{$#1$}{$\times$} \, \phantom{10^{#3}}}

\newcommand{\AMRtolerance}{\varepsilon_\text{lte}}

\newcommand{\ErrorToleranceEffective}{\varepsilon_{\text{lte,eff}}}

\newcommand{\FPepsilon}{\varepsilon_\text{fp}}

\newcommand{\FPlimit}[1]{$\llbrack #1 \rrbrack$}
\newcommand{\FPlimitPerGridPoint}{\FPlimit{1}}
\newcommand{\FPlimitLTEEstimate}{\FPlimit{2}}

\newcommand{\shadecell}{\cellcolor[gray]{0.60}}
\newcommand{\shaderow}{\rowcolor[gray]{0.60}[1.75\tabcolsep][1.75\tabcolsep]}

\begin{document}
\title{Highly accurate and efficient self-force computation
       using time-domain methods:
       Error estimates, validation, and optimization}
\author{Jonathan Thornburg}
\affiliation{Department of Astronomy,
	     Indiana University,
	     Bloomington, Indiana, USA}
\email{jthorn@astro.indiana.edu}
%
%
\date{$ $Id: hasf.tex,v 1.356 2010/06/18 19:35:16 jonathan Exp $ $}


\begin{abstract}
If a small ``particle'' of mass $\mu M$ (with $\mu \ll 1$) orbits
a Schwarzschild or Kerr black hole of mass $M$, the particle is subject
to an $\O(\mu)$ radiation-reaction ``self-force''.  Here I argue that
it's valuable to compute this self-force highly accurately
(relative error of $\ltsim 10^{-6}$) and efficiently, and I describe
techniques for doing this and for obtaining and validating error
estimates for the computation.  I use an adaptive-mesh-refinement (AMR)
time-domain numerical integration of the perturbation equations in
the Barack-Ori mode-sum regularization formalism; this is efficient,
yet allows easy generalization to arbitrary particle orbits.  I focus
on the model problem of a scalar particle in a circular geodesic orbit
in Schwarzschild spacetime.

The mode-sum formalism gives the self-force as an infinite sum of
regularized spherical-harmonic modes $\sum_{\ell=0}^\infty F_{\ell,\reg}$,
with $F_{\ell,\reg}$ (and an ``internal'' error estimate) computed
numerically for $\ell \ltsim 30$ and estimated for larger~$\ell$ by
fitting an asymptotic ``tail'' series.  Here I validate the internal
error estimates for the individual $F_{\ell,\reg}$ using a large set
of numerical self-force computations of widely-varying accuracies.
I present numerical evidence that the actual numerical errors in
$F_{\ell,\reg}$ for different~$\ell$ are at most weakly correlated,
so the usual statistical error estimates are valid for computing the
self-force.  I show that the tail fit is numerically ill-conditioned,
but this can be mostly alleviated by renormalizing the basis functions
to have similar magnitudes.

Using AMR, fixed mesh refinement, and extended-precision floating-point
arithmetic, I obtain the (contravariant) radial component of the self-force
for a particle in a circular geodesic orbit of areal radius $r = 10M$
to within $1$~ppm relative error, as estimated both by internal error
estimates and by comparison with previously-published frequency-domain
calculations.
\end{abstract}



\maketitle


\begin{quote}
\textit{This paper is dedicated to the memory of Thomas Radke,
my late friend, colleague, and partner in many computational
adventures.}
\end{quote}


\section{Introduction}
\label{sect-introduction}

Consider a small ``particle'' of mass $\mu M$ (with $\mu \ll 1$)
moving freely in an asymptotically-flat background spacetime, say
for definiteness Schwarzschild or Kerr spacetime of mass $M$.  This
system emits gravitational waves (GWs), and there is a corresponding
radiation-reaction influence on the particle's motion.  Self-consistently
calculating this motion and the emitted gravitational radiation is
a long-standing research question, and is interesting both as an
abstract problem in general relativity and as an essential prerequisite
for the success of the proposed Laser Interferometer Space Array
(LISA) space-based gravitational radiation detector.  A typical LISA
extreme mass ratio inspiral (EMRI) source is expected to comprise
a stellar-mass black hole or neutron star (the ``particle'') orbiting
a supermassive black hole with $M \sim 10^6 M_\sun$,
\footnote{
	 Here $M_\sun$ denotes the solar mass.
	 }
{} so that $\mu \sim 10^{-5}$ to~$10^{-6}$; the particle's orbit
will typically be both inclined (with respect to the supermassive
black hole's equatorial plane defined by its spin) and
moderately-to-highly eccentric.  LISA is expected to observe many
such systems, some of them at quite high signal/noise ratios once
the raw data stream is matched-filtered against appropriate waveform
templates~(\cite{Gair-etal-2004:LISA-EMRI-event-rates,Barack-Cutler-2004,
Amaro-Seoan-etal-2007:LISA-IMRI-and-EMRI-review,
Gair-2009:LISA-EMRI-event-rates}; see
section~\ref{sect-importance-of-highly-accurate-efficient/accuracy/LISA}
for further discussion).

The particle's orbit may be highly relativistic,
so post-Newtonian methods (see, for
example,~\cite[section~6.10]{Damour-in-Hawking-Israel-1987};
\cite{Blanchet-2006-living-review,Futamase-Itoh-2007:PN-review,
Blanchet-2009:PN-review,Schaefer-2009:PN-review} and references therein)
may not be accurate for this problem.  Since the timescale for
radiation reaction to shrink the orbit is very long ($\sim \mu^{-1} M$)
while the required resolution near the particle is very high
($\sim \mu M$), full numerical-relativity methods (see, for
example,~\cite{Pretorius-2007:2BH-review,Hannam-etal-2009:Samurai-project,
Hannam-2009:2BH-review,
Hannam-Hawke-2009:2BH-in-era-of-Einstein-telescope-review,
Campanelli-etal-2010:2BH-numrel-review} and references therein)
are prohibitively expensive for this problem.
\footnote{
	 A number of researchers have attempted to develop
	 special numerical-relativity methods to make such
	 simulations practical, at least for systems with
	 ``intermediate'' mass ratios $\mu \sim 10^{-3}$.
	 Although promising initial results have been obtained
(see, for example,
\cite{Bishop-etal-2003,Bishop-etal-2005,
Sopuerta-etal-2006,Sopuerta-Laguna-2006,
Lousto-etal-2010:intermediate-mass-2BH-numrel-Lazarus}),
	 it has not (yet) been possible to perform numerical
	 evolutions lasting for radiation-reaction time
	 scales.
	 }

Instead, it's appropriate to use black hole perturbation theory,
treating the particle as an $\O(\mu)$ perturbation on the background
Schwarzschild or Kerr spacetime.  A self-consistent calculation of
the emitted gravitational radiation requires knowledge of the metric
perturbation induced by the particle up to and including $\O(\mu^2)$
terms (\cite[section~5.5.6]{Poisson-2004-living-review};
\cite[section~11.1]{Detweiler-2005}; \cite{Burko-2003,Burko-2005}).
The theoretical formalism for such calculations is not yet fully
developed;
\footnote{
	 See, for
example,~\cite{Rosenthal-2005:2nd-order-scalar-regularization,
Rosenthal-2005:2nd-order-grav-regularization,
Rosenthal-2006:2nd-order-grav-perturbation,
Rosenthal-2006:2nd-order-grav-self-force}
	 for recent work towards $\O(\mu^2)$ calculations.
	 }
{} here I present calculations only for the $\O(\mu)$~self force.

Building on the early work of DeWitt and Brehme~\cite{DeWitt-Brehme-1960}
(with a correction by Hobbs~\cite{Hobbs-1968}),
\footnote{
	 Another significant early work is that of
	 Gal'tsov~\cite{Galtsov-1982}, but this approach has
	 serious causality difficulties: in a curved spacetime
	 it gives the self-force at a specified time in terms
	 of the \emph{future} evolution of the particle.
	 }
{} the $\O(\mu)$ ``MiSaTaQuWa'' equations of motion for a gravitational
point particle in a (strong-field) curved spacetime were first derived
by Mino, Sasaki, and Tanaka~\cite{Mino-Sasaki-Tanaka-1997} and
Quinn and Wald~\cite{Quinn-Wald-1997} (see also Detweiler's
analysis~\cite{Detweiler-2001:radiation-reaction-and-self-force}),
and have recently been rederived in a more rigorous manner by
Gralla and Wald~\cite{Gralla-Wald-2008}.
\footnote{
\label{footnote-GHW-proof-of-EM-self-force}
	 Gralla, Harte, and Wald~\cite{Gralla-Harte-Wald-2009}
	 have also recently obtained a rigorous derivation of
	 the electromagnetic self-force in a curved spacetime.
	 }
{}  See \cite{Poisson-2004-living-review,Detweiler-2005,
Barack-2009:self-force-review,Detweiler-2009:self-force-review,
Poisson-2009:self-force-review} for general reviews of the
self-force problem.

The particle's motion may be modelled as either (i) non-geodesic motion
in the background Schwarzschild/Kerr spacetime under the influence of
a radiation-reaction ``self-force'', or (ii) geodesic motion in a
perturbed spacetime.  These two perspectives are
equivalent~\cite{Sago-Barack-Detweiler-2008}; in this work I use~(i).
The MiSaTaQuWa equations then give the self-force in terms of (the
gradient of) the metric perturbation due to the particle, which must
be computed using black-hole perturbation theory.

The computation of the metric perturbation due to a point particle
is particularly difficult because the ``perturbation'' is formally
infinite at the particle.  A practical ``mode-sum'' scheme to
regularize the metric perturbation was developed by
Barack and Ori~\cite{Barack-Ori-2000,Barack-2000,Barack-etal-2002,
Barack-Ori-2002,Barack-Ori-2003}, and in slightly different forms
by Detweiler, Messaritaki, and
Whiting~\cite{Detweiler-Whiting-2003,Detweiler-Messaritaki-Whiting-2003}
and Haas and Poisson~\cite{Haas-Poisson-2006}.
Here I follow the Barack-Ori ``$\ell$-mode'' regularization
(described in detail in~\cite{Barack-Ori-2003} and summarized in
section~\ref{sect-Barack-Ori}).  This is based on a spherical-harmonic
decomposition of the metric perturbation, allowing the 4-vector
self-force $F^a$ to be written as an infinite sum of regularized
modes $F^a = \sum_{\ell=0}^\infty F^a_{\ell,\reg}$.  Each regularized
mode $F^a_{\ell,\reg}$ is calculated by solving a set of linear
partial differential equations (PDEs), computing certain derivatives
of the PDE solutions along the particle worldline, and finally
subtracting certain analytically-known regularization coefficients.

Depending on how the PDEs are solved, there are two broad classes of
self-force computations within the mode-sum regularization framework:
frequency-domain and time-domain.  Frequency-domain computations
involve a Fourier transform of each mode's PDEs in time, reducing
the numerical computation to the solution of a set of ordinary
differential equations (ODEs) for each mode (see, for example,
\cite{Detweiler-Messaritaki-Whiting-2003}).  Frequency-domain
computations are typically very efficient and accurate for
circular or near-circular particle orbits,
\footnote{
	 As a notable example of this accuracy,
	 Blanchet~\etal{}~\cite{Blanchet-etal-2009:cmp-3PN-with-self-force}
	 have recently computed the gravitational self-force
	 for circular geodesic orbits in Schwarzschild
	 spacetime to a relative accuracy of approximately
	 one part in $10^{13}$.
 	 }
{} but degrade rapidly in efficiency with increasing eccentricity
of the particle's orbit, becoming impractical for highly eccentric
orbits~\cite{Glampedakis-Kennefick-2002,Barack-Lousto-2005}.
\footnote{
	 Barack, Ori, and Sago~\cite{Barack-Ori-Sago-2008}
	 have recently found an elegant solution for some
	 other limitations which had previously affected
	 frequency-domain calculations.
	 }
{}  In contrast, time-domain computations involve a direct numerical
integration of each mode's PDEs, and have traditionally been somewhat
less efficient and accurate than frequency-domain computations.
However, time-domain computations can accommodate arbitrary particle
orbits with only minor penalties in performance and accuracy
(\cite{Barton-etal:cmp-EMRI-frequency-vs-time-domain-methods}),
and some complications in the numerical schemes
(see, for example, \cite{Haas-2007,Barack-Sago-2010}).

In this work I use the time-domain approach, using an adaptive mesh
refinement (AMR) code with 4th~order finite
differencing~\cite{Thornburg-2009:characteristic-AMR} to solve each
mode's PDEs very accurately and efficiently.  To simplify the boundary
treatment, I use a characteristic (double-null) evolution scheme.
I restrict consideration to the model problem of computing the self-force
on a scalar particle moving in Schwarzschild spacetime.  This is a
widely-used test problem in the field of self-force calculations,
with past numerical computations
including~\cite{Barack-Burko-2000,Barack-2000,Burko-2000a,Burko-2000b,
Detweiler-Messaritaki-Whiting-2003,Diaz-Rivera-etal-2004,
Haas-Poisson-2006,Haas-2007,Vega-Detweiler-2008:self-force-regularization,
Vega-etal-2009:self-force-3+1-primer,
Canizares-Sopuerta-2009a,Canizares-Sopuerta-2009b}.
\footnote{
	 The electromagnetic self-force (a more complicated
	 ``toy model'' by virtue of the nontrivial gauge
	 freedom) has been studied by~\cite{Keidl-Friedman-Wiseman-2007}.
	 (Note also the recent work described in
	 footnote~\ref{footnote-GHW-proof-of-EM-self-force}.)
	 The gravitational self-force has been studied by
	 numerous authors,
including~\cite{Barack-Lousto-2002,Barack-Lousto-2005,
Keidl-Friedman-Wiseman-2007,Barack-Sago-2007,Barack-Sago-2009,
Detweiler-2008:grav-self-force-for-circular-Schw-orbits,
Vega-Detweiler-2008:self-force-regularization,
Blanchet-etal-2009:cmp-3PN-with-self-force,Barack-Sago-2010}.
	 }
$^,$
\footnote{
	 Warburton and Barack~\cite{Warburton-Barack-2010}
	 have recently reported results for the self-force
	 on a scalar charge in a circular equatorial geodesic
	 orbit in \emph{Kerr} spacetime.
	 }
{}  For the numerical computations presented here, I further restrict
consideration to the computation of the radial component of the
self-force for a scalar particle in a circular geodesic orbit about
the Schwarzschild black hole.  However, I also simulate the accuracy
to be expected when similar methods are applied to generic non-circular
particle orbits.

The basic mode-sum technique for self-force computation discussed
here is already well-known.  The main new results in this paper
concern
(a) the (small) extension of these techniques to accommodate the use
of characteristic AMR for the numerical integrations,
(b) the error estimates for such a self-force computation,
(c) the validation of these error estimates using a large set of
numerical computations of widely-varying accuracies, 
(d) the tail fit's ill-conditioning,
(e) the cost/accuracy tradeoffs for the computation, and
(f) the demonstration of consistency at levels of $\sim\! 0.1$ parts
per million (ppm) relative error between the time-domain self-force
computations presented here and the highly-accurate frequency-domain
computations of Detweiler, Messaritaki, and
Whiting~\cite{Detweiler-Messaritaki-Whiting-2003}.

The remainder of this paper is organized as follows:
Section~\ref{sect-introduction/notation} outlines the notation used
in this paper.
Section~\ref{sect-importance-of-highly-accurate-efficient} discusses
the scientific importance of highly accurate and efficient self-force
computations.
Section~\ref{sect-Barack-Ori} outlines the Barack-Ori mode-sum
regularization procedure for self-force computations.
Section~\ref{sect-numerical-methods} outlines the numerical methods
I use for the self-force calculation and its error estimates.
Section~\ref{sect-results} presents my numerical results.
Section~\ref{sect-conclusions} presents conclusions and directions
for further research.


\subsection{Notation}
\label{sect-introduction/notation}

I generally follow the sign and notation conventions of Wald~\cite{Wald84},
with $G = c = 1$ units and a $(-,+,+,+)$ metric signature.  I use the
Penrose abstract-index notation, with Latin indices $ab$ running over
spacetime coordinates.  $g$ is the determinant of the 4-metric
and $\del_a$ the associated covariant derivative operator.
$\boxop \eqdef \del_a \del^a$ is the 4-dimensional wave operator.
$\| \cdot \|_\rms$ is the root-mean-square norm on $\Re^n$,
$\big\| \{x_k\} \big\|_\rms \eqdef \sqrt{ \left( \sum_k x_k^2 \right)/n}$.


\section{The Importance of Highly Accurate and Efficient
	 Self-Force Calculations}
\label{sect-importance-of-highly-accurate-efficient}

In this section I outline several different lines of argument
suggesting that it's scientifically valuable to compute the EMRI
self-force highly accurately and efficiently.


\subsection{The Importance of High Accuracy}
\label{sect-importance-of-highly-accurate-efficient/accuracy}


\subsubsection{LISA}
\label{sect-importance-of-highly-accurate-efficient/accuracy/LISA}

A major part of the motivation for self-force calculations comes
from their planned application to EMRI data analysis for LISA.  In
the words of
Amaro-Seoane~\etal~\cite[section~3.1]{Amaro-Seoan-etal-2007:LISA-IMRI-and-EMRI-review},
\begin{quote}
A typical EMRI signal will have an instantaneous amplitude an order
of magnitude below the LISA's instrumental noise and (at low frequencies)
as many as several orders of magnitude below the gravitational wave
foreground from Galactic compact binaries.  This makes detection a
rather difficult problem.  However, the signals are very long lived,
and will be observed over more than $10^5$~cycles, which in principle
allows the signal-to-noise ratio (SNR) to be built up over time using
matched filtering.
\end{quote}

Matched filtering of the entire years-long LISA data stream would be
impractically expensive for \emph{detecting} EMRIs with hitherto-unknown
parameters~\cite[section~3]{Gair-etal-2004:LISA-EMRI-event-rates}.
However, once EMRIs have been detected by more economical search
algorithms~(\cite[section~3.1]{Amaro-Seoan-etal-2007:LISA-IMRI-and-EMRI-review};
\cite{Porter-2009:LISA-data-analysis-overview}),
precision modelling and matched filtering of the full LISA data
stream become practical, allowing accurate measurements of the
EMRI parameters, tests of general relativity, and other valuable
astrophysical measurements
(see, for example, \cite{Ryan-1995,Ryan-1997,Barack-Cutler-2004,
Hopman-2006:astrophysics-of-EMRI-sources, Glampedakis-Babak-2006,
Barack-Cutler-2007};
\cite[sections~4 and~5]{Amaro-Seoan-etal-2007:LISA-IMRI-and-EMRI-review};
\cite{Miller-etal:probing-stellar-dynamics-in-galactic-nuclei}).

Gair~\cite{Gair-2009:LISA-EMRI-event-rates} has recently updated past
calculations~\cite{Gair-etal-2004:LISA-EMRI-event-rates} of LISA EMRI
event rates and has calculated the redshift~$z$ of the closest~($z_{\min}$)
and most distant~($z_{\max}$) EMRIs that LISA is likely to detect
under a range of assumptions about the LISA mission duration and
hardware reliability, the supermassive black hole's spin, and the
EMRI rate per galaxy.  For these calculations,
Gair~\cite{Gair-2009:LISA-EMRI-event-rates} assumed a detection
threshold of $\rho_\thresh = 30$, where $\rho$~is the EMRI signal-to-noise
ratio after matched filtering.  That is, $z_{\max}$ is the redshift at
which the strongest expected LISA EMRI will have a signal-to-noise ratio
(after matched filtering) of $\rho_\thresh$.  Neglecting cosmological
spacetime curvature, the signal-to-noise ratio for a given source
scales inversely with~$z$, so (neglecting Malmquist bias)
\footnote{
	 Malmquist bias is a selection effect in a brightness-limited
	 sample: nearby objects are included in the sample regardless
	 of their intrinsic luminosity, but intrinsically-faint
	 distant objects fall below the sample's minimum-brightness
	 threshold and are thus omitted from the sample.  The result
	 is that the mean intrinsic luminosity of sample objects
	 increases with distance~\cite{Malmquist-1920,
Teerikorpi-1997:Malmquist-bias-etal-review}.  In the present
	 context, this results in the the $z_{\max}$~EMRI being
	 intrinsically brighter than the $z_{\min}$~EMRI, which
	 somewhat reduces~$\rho_{\max}$.
	 }
{} the signal-to-noise ratio of the \emph{closest} LISA EMRI (which
I take as an approximation to the \emph{strongest} LISA EMRI) is thus
$\rho_{\max} \approx (z_{\max}/z_{\min}) \rho_\thresh$.
{}  Table~\ref{tab-LISA-EMRI-signal/noise} gives the resulting $\rho_{\max}$
for each of Gair's~\cite[table~4]{Gair-2009:LISA-EMRI-event-rates}
LISA-performance and astrophysics assumptions.  The $\rho_{\max}$
values range from ${\sim}\, 20$ to as high as ${\sim}\, 2000$.

\begin{table}[bp]
\begin{center}
Signal-to-Noise Ratios of the Strongest LISA EMRIs			\\
\begin{ruledtabular}
\begin{tabular}{lcrccrccrc}
		&&		&& \multicolumn{6}{c}{LISA Performance}	\\
\cline{5-10}
$a$		& \multicolumn{3}{c}{$\mathcal{R}^\text{BH}_\text{MW}$~($\text{Gyr}^{-1}$)}
				& \multicolumn{3}{c}{5yr,2chan}
						& \multicolumn{3}{c}{2yr,1chan}
									\\
\hline 
0\P{.0}
	\quad
	$\left\{
	 \begin{tabular}[c]{c}
	 \hbox{}	\\
	 \hbox{}	\\
	 \hbox{}	
	 \end{tabular}
	 \right.$
		&& \begin{tabular}[c]{r}
		     4		\\
		    40		\\
		   400		
		   \end{tabular}
				&&& \begin{tabular}[c]{r}
				   140		\\
				   460		\\
				  1300		
				  \end{tabular}
						&&& \begin{tabular}[c]{r}
						    16\rlap{*}	\\
						   180		\\
						   490		
						  \end{tabular}
								&	\\[5ex]
0.5
	\quad
	$\left\{
	 \begin{tabular}[c]{c}
	 \hbox{}	\\
	 \hbox{}	\\
	 \hbox{}	
	 \end{tabular}
	 \right.$
		&& \begin{tabular}[c]{r}
		     4		\\
		    40		\\
		   400		
		   \end{tabular}
				&&& \begin{tabular}[c]{r}
				   160		\\
				   560		\\
				  1500		
				  \end{tabular}
						&&& \begin{tabular}[c]{r}
						    20\rlap{*}	\\
						   210		\\
						   590		
						  \end{tabular}
								&	\\[5ex]
0.9
	\quad
	$\left\{
	 \begin{tabular}[c]{c}
	 \hbox{}	\\
	 \hbox{}	\\
	 \hbox{}	
	 \end{tabular}
	 \right.$
		&& \begin{tabular}[c]{r}
		     4		\\
		    40		\\
		   400		
		   \end{tabular}
				&&& \begin{tabular}[c]{r}
				   240		\\
				   780		\\
				  2100		
				  \end{tabular}
						&&& \begin{tabular}[c]{r}
						    46\rlap{*}	\\
						   330		\\
						   860		
						  \end{tabular}
								&	\\[1ex]
\end{tabular}
\end{ruledtabular}
\end{center}
\caption[Signal-to-Noise Ratios of the Strongest LISA EMRIs]
	{
	This table shows the estimated signal-to-noise ratio
	after matched filtering, $\rho_{\max}$, of the closest
	(approximately the strongest) LISA EMRI sources.  $a$~is
	the dimensionless spin of the EMRI's central supermassive
	black hole, $\mathcal{R}^\text{BH}_\text{MW}$~is the
	EMRI rate for the Milky Way galaxy, and ``5yr,2d'' and
	``2yr,1d'' refer to different assumptions about the
	LISA mission lifetime (5 versus 2~years) and hardware
	reliability (2chan = full configuration with 2~independent
	low-frequency interferometer channels available;
	1chan = degraded configuration with only 1~independent
	low-frequency interferometer channel available).
	``*'' marks values which are very uncertain due to
	small-$N$ statistics in Gair's
	simulations~\cite[table~4]{Gair-2009:LISA-EMRI-event-rates}.
	}
\label{tab-LISA-EMRI-signal/noise}
\end{table}

In order to achieve these high signal-to-noise ratios, LISA will
require matched filtering against accurate EMRI GW templates.  In
order to keep parameter-estimation errors
\footnote{
	 These parameters might be those characterizing
	 the EMRI system itself, those characterizing the
	 deviation of the supermassive-body spacetime from
	 the Kerr metric, or those for other tests of
	 general relativity
(see, for example, \cite{Ryan-1995,Ryan-1997,Barack-Cutler-2004,
Glampedakis-Babak-2006,Barack-Cutler-2007};
\cite[sections~4 and~5]{Amaro-Seoan-etal-2007:LISA-IMRI-and-EMRI-review};
\cite[section~5]{Psaltis-2008}; \cite{Vigeland-Hughes-2010:bumpy-BHs}).
	 }
{} due to template inaccuracy below those due to statistical noise,
the (template) EMRI GW phase must be modelled to an accuracy of
$\Delta\phi \ltsim C/\rho_{\max}$ radians over the LISA mission
lifespan, where the ``degeneracy factor''~$C$ depends on the level
of degeneracy between the different parameters for the particular
analysis being done.  $C$ is often estimated via the Fisher-matrix
formalism (see, for example,
\cite{Finn-1992,Cutler-Flanagan-1994,Jaranowski-Krolak-2005,
Cutler-Vallisneri-2007,Vallisneri-2008:Fisher-matrix-use-and-abuse,
Huerta-Gair-2009} and references therein).
\footnote{
Lindblom~\etal~\cite{Lindblom-Owen-Brown-2008:model-waveform-standards,
Lindblom-2009a:optimal-calibration-accuracy-for-GW-detectors,
Lindblom-2009b:use-and-abuse-of-model-waveform-standards}
	 have carefully quantified a similar line of
	 reasoning for the case of comparable-mass black
	 hole binaries.
	 }

LISA will observe an EMRI for $N \sim 2\pi \,{\cdot}\, 10^5$~radians
of GW phase (see, for example, \cite[table~I]{Huerta-Gair-2009}), so
the accuracy tolerance for the allowable GW phase error corresponds to
a relative tolerance $\Delta\dot{\phi}/\dot{\phi} \ltsim C/(N \rho_{\max})$
for the instantaneous GW frequency.  Table~\ref{tab-LISA-EMRI-phase-accuracy}
gives these tolerances for degeneracy parameters $C = 1$~(very optimistic),
$C = 30$~(reasonable for many tests-of-GR analyses), and
$C = 1000$~(somewhat pessimistic).

\begin{table}[bp]
\begin{center}
Gravitational-Wave Phase						\\
Error Tolerance $\Delta\phi$ (radians)					\\
\begin{ruledtabular}
\begin{tabular}{l@{\extracolsep{0.5em}}ddd}
			& \multicolumn{1}{c}{$C = 1$}
			& \multicolumn{1}{c}{$C = 30$}
			& \multicolumn{1}{c}{$C = 1000$}		\\
\cline{2-2}\cline{3-3}\cline{4-4}					\\[-2ex]
$\rho_{\max} = 30$	& 0.03		& 1		& 30		\\
$\rho_{\max} = 300$	& 0.003		& 0.1		&  3		\\
$\rho_{\max} = 2000$	& 0.0005	& 0.015		&  0.5		\\
\end{tabular}
\end{ruledtabular}
\end{center}
\medskip
\begin{center}
Instantaneous Gravitational-Wave Frequency				\\
Fractional Error Tolerance $\Delta\dot{\phi}/\dot{\phi}$		\\
\begin{ruledtabular}
\begin{tabular}{l@{\extracolsep{0.5em}}lll}				\\
			& \multicolumn{1}{c}{$C = 1$}
			& \multicolumn{1}{c}{$C = 30$}
			& \multicolumn{1}{c}{$C = 1000$}		\\
\cline{2-2}\cline{3-3}\cline{4-4}					\\[-2ex]
$\rho_{\max} = 30$	& $\SciNum{5}{-8}$
			& $\SciNum{2}{-6}$
			& $\SciNum{5}{-5}$				\\
$\rho_{\max} = 300$	& $\SciNum{5}{-9}$
			& $\SciNum{2}{-7}$
			& $\SciNum{5}{-6}$				\\
$\rho_{\max} = 2000$	& $\SciNum{8}{-10}$
			& $\SciNum{2}{-8}$
			& $\SciNum{8}{-7}$				\\
\end{tabular}
\end{ruledtabular}
\end{center}
\caption[LISA EMRI Gravitational-Wave Error Tolerances]
	{
	This table shows the maximum errors allowed in an EMRI
	gravitational-wave template so that the resulting
	parameter-estimation errors for the strongest expected
	LISA EMRI do not exceed the statistical errors due to
	LISA's instrumental and confusion noise levels, given
	various combinations of the EMRI signal-to-noise ratio
	$\rho_{\max}$ (after matched filtering) and the
	parameter degeneracy factor~$C$.
	The error tolerances are expressed alternatively as a total
	phase error $\Delta\phi$~(radians), or as a (dimensionless)
	relative error in the instantaneous gravitational-wave
	frequency, $\Delta\dot{\phi}/\dot{\phi}$.
	}
\label{tab-LISA-EMRI-phase-accuracy}
\end{table}

These GW error tolerances can be related to the required accuracy in a
self-force computation using the results of
Huerta and Gair~\cite[table~I]{Huerta-Gair-2009},
who estimate the effects of various $\O(\mu^2)$ self-force effects
-- that is, $\O(\mu^2/\mu) \sim 10^{-5}$ fractional changes in the
overall $\O(\mu)$~self force -- on an EMRI's GW phase.  They find
that $\O(\mu^2)$~effects change the cumulative EMRI GW phase by
$\sim 3$~orbits (20~radians) over the $\sim 10^5$-orbit LISA observation
span.  Equivalently, a 1~part per million (ppm) fractional change in the
overall $\O(\mu)$~self force changes the cumulative EMRI GW phase by
approximately $0.3$~orbits ($2$~radians).  The LISA EMRI phase error
tolerances given in table~\ref{tab-LISA-EMRI-phase-accuracy} thus
correspond to the self-force accuracy tolerances given in
table~\ref{tab-LISA-EMRI-self-force-accuracy}.  It's clear that
self-force computations accurate to between roughly one part per
million and one part per billion are required to avoid degrading
the parameter-estimation accuracy for the strongest LISA EMRIs.

\begin{table}[bp]
\begin{center}
Self-Force Relative Error Tolerance				       \\[0.5ex]
\begin{ruledtabular}
\begin{tabular}{l@{\extracolsep{0.5em}}lll}
			& \multicolumn{1}{c}{$C = 1$}
			& \multicolumn{1}{c}{$C = 30$}
			& \multicolumn{1}{c}{$C = 1000$}		\\
\cline{2-2}\cline{3-3}\cline{4-4}					\\[-2ex]
$\rho_{\max} = 30$	& $\SciNum{2}{-8}$
			& $\SciNum{5}{-7}$
			& $\SciNum{2}{-5}$				\\
$\rho_{\max} = 300$	& $\SciNum{2}{-9}$
			& $\SciNum{5}{-8}$
			& $\SciNum{2}{-6}$				\\
$\rho_{\max} = 2000$	& $\SciNum{3}{-10}$
			& $\SciNum{8}{-9}$
			& $\SciNum{3}{-7}$				\\
\end{tabular}
\end{ruledtabular}
\end{center}
\caption[LISA EMRI Self-Force Error Tolerances]
	{
	This table shows the maximum relative errors allowed
	in an EMRI self-force computation so that the resulting
	parameter-estimation errors for the strongest expected
	LISA EMRI do not exceed the statistical errors due to
	LISA's instrumental and confusion noise levels, given
	different combinations of the EMRI signal-to-noise
	ratio~$\rho_{\max}$ (after matched filtering) and
	the parameter degeneracy factor~$C$.
	}
\label{tab-LISA-EMRI-self-force-accuracy}
\end{table}


\subsubsection{Self-Force Calculations}
\label{sect-importance-of-highly-accurate-efficient/accuracy/self-force}

As noted earlier, computing EMRI GW waveforms in a fully self-consistent
manner requires calculating the metric perturbation induced by the
particle -- and the corresponding self-force -- up to and including
at least $\O(\mu^2)$ terms (\cite[section~5.5.6]{Poisson-2004-living-review},
\cite[section~11.1]{Detweiler-2005}, \cite{Burko-2003,Burko-2005}),
but the theoretical formalism for doing this isn't fully developed
yet.

However, in the near future some $\O(\mu^2)$~effects are likely to
be explored with ``orbit correction''
calculations~\cite[section~7]{Gralla-Wald-2008}), where the
$\O(\mu)$~self force is used to calculate the time evolution of
the orbit parameters.  In order to reliably distinguish true
$\O(\mu^2)$~effects due to the orbit correction from numerical errors
in the $\O(\mu)$~self force, the $\O(\mu)$~self force needs to be
calculated with a relative error $\ll \mu \sim 10^{-5}$.

This same argument should continue to hold once (if) future self-force
calculations are able to include all $\O(\mu^2)$~effects and compute
GW waveforms in a fully self-consistent manner.

Rosenthal's work towards $\O(\mu^2)$~self-force
calculations~\cite{Rosenthal-2005:2nd-order-scalar-regularization,
Rosenthal-2005:2nd-order-grav-regularization,
Rosenthal-2006:2nd-order-grav-perturbation,
Rosenthal-2006:2nd-order-grav-self-force}
suggests that the $\O(\mu)$~metric perturbation will be needed
to high accuracy as an input into the $\O(\mu^2)$~calculations.

Highly accurate self-force calculations are also valuable for helping
to calibrate and constrain various terms in post-Newtonian expansions
multiple-body systems (see, for
example,~\cite{Damour-2009:Schw-grav-self-force-and-EOB,
Detweiler-2008:grav-self-force-for-circular-Schw-orbits,
Blanchet-etal-2009:cmp-3PN-with-self-force} and references therein).

Finally, highly accurate calculations of the $\O(\mu)$~self force are
valuable as a test case for the intricate theory and computations
involved.  For the calculations reported here I use time-domain
integrations of the metric-perturbation equations in the Barack-Ori
mode-sum formalism.  In contrast, the most accurate published calculation
of the self-force for this case, that of Detweiler, Messaritaki, and
Whiting~\cite{Detweiler-Messaritaki-Whiting-2003}, uses a
frequency-domain approach with completely different numerical methods.
Precisely because the two calculations are structured so differently,
a verification of their agreement to high precision serves as a useful
check on both techniques and their respective theoretical formalisms.
\footnote{
\label{footnote-cmp-different-calculations}
	 Sago, Barack, and Detweiler~\cite{Sago-Barack-Detweiler-2008}
	 and Barack and Sago~\cite{Barack-Sago-2010} have
	 previously compared time- and frequency-domain
	 self-force calculations.  Comparisons of self-force
	 calculations with post-Newtonian expansions
(see, for example, \cite{Damour-2009:Schw-grav-self-force-and-EOB,
Detweiler-2008:grav-self-force-for-circular-Schw-orbits,
Blanchet-etal-2009:cmp-3PN-with-self-force} and references therein)
	 also implicitly check the correctness of both.
	 }


\subsection{The Importance of High Efficiency}
\label{sect-importance-of-highly-accurate-efficient/efficiency}

The precision modelling and matched filtering of a single already-detected
EMRI is essentially a many-parameter nonlinear least-squares fitting
process, and thus requires generating many trial waveforms.  Moreover,
this process should be repeated for each strong EMRI source, of which
there will likely be many~\cite{Gair-etal-2004:LISA-EMRI-event-rates}.

With current methods, a single EMRI self-force calculation takes
between one-half and one cpu-week at the $10^{-4}$ relative-error
level~\cite[section~III.E]{Barack-Sago-2010}.  This is already
unpleasantly slow, and raising the accuracy to the
$\ltsim 10^{-6}$~relative-error level will slow the computation by
another factor of $\sim 10$,
\footnote{
	 Like my code, Barack and Sago's~\cite{Barack-Sago-2010}
	 code uses globally 4th~order finite differencing,
	 so a $\times 100$~accuracy improvement requires a
	 $\times \sqrt{10}$~increase in resolution, which
	 costs a factor of~$10$ in CPU time for a $1{+}1$-dimensional
	 evolution.
	 }
{} although parallelization should be easy.

Unfortunately, actual EMRI \emph{waveform} calculations will likely
be much \emph{slower} than self-force calculations.  For example,
an orbit-correction calculation essentially requires time-integrating
a set of coupled ODEs for the orbital-parameter evolution on
radiation-reaction and longer timescales, with the ODEs' right-hand-side
functions being given by a self-force
computation~\cite[section~7]{Gralla-Wald-2008}).
Even the most efficient ODE-integration schemes~\cite{Gear-1981:ODEs-review}
will require evaluating the right-hand-side functions (i.e., computing
the self-force for some specified intermediate orbit) hundreds of times
in the course of a single orbit-correction calculation, so the need
for the highest possible efficiency in the self-force computation
is clear.


\section{Self-Force Calculation via the Barack-Ori Mode-Sum Regularization}
\label{sect-Barack-Ori}

In this section I briefly outline the Barack-Ori mode-sum regularization
procedure for computing the self force, for the special case of a
scalar particle in a circular geodesic orbit in Schwarzschild
spacetime.  A more detailed account can be found in the original
works by Barack and Ori~\cite{Barack-Ori-2000,Barack-2000,
Barack-etal-2002,Barack-Ori-2002,Barack-Ori-2003}.  I defer most
discussion of numerical methods for this calculation to
section~\ref{sect-numerical-methods}.


\subsection{Schwarzschild spacetime}
\label{sect-Barack-Ori/Schw}

Consider Schwarzschild spacetime of mass~$M$, and introduce ingoing
and outgoing null coordinates $u$ and $v$ respectively, so the line
element is
\begin{equation}
ds^2 = - f(r) \, du \, dv + r^2 (d\theta^2 + \sin^2 \theta \, d\varphi^2)
								      \,\text{,}
\end{equation}
where $r$ is the usual areal radial coordinate, $f(r) \eqdef 1 - 2M/r$,
and $(\theta,\varphi)$ are the usual polar spherical angular coordinates
on a 2-sphere of constant~$r$.  It's also useful to define the
Schwarzschild time coordinate $t_\Schw = \thalf (v+u)$ and the
``tortise'' radial coordinate
\begin{equation}
r_* = \thalf (v-u) = r + 2M \log \left| \frac{r}{2M} - 1 \right|
								      \,\text{.}
\end{equation}

It's convenient to define the specific energy~$\E$, specific angular
momentum~$\L$, and orbital frequency~$\omega$ of a test particle
in a circular geodesic orbit at the areal radius $r$,
\begin{eqnarray}
\E(r)	& = &	\frac{f(r)}{\sqrt{1 - 3M/r}}				\\
\L(r)	& = &	\frac{\sqrt{Mr}}{\sqrt{1 - 3M/r}}			\\
\omega(r)
	& = &	\sqrt{\frac{M}{r^3}}
								      \,\text{,}
\end{eqnarray}


\subsection{The Scalar Field}
\label{sect-Barack-Ori/scalar-field}

I take the real scalar field~$\Phi$ to satisfy the equation
\begin{equation}
\boxop \Phi
	= - 4 \pi q
	  \int_{-\infty}^\infty
	  \frac{\delta^4 \bigl(x^a - x^a_p(\tau) \bigr)}{\sqrt{-g}}
	  \, d\tau
								      \,\text{,}
\end{equation}
where $q$ is the particle's scalar charge and
$\tau$ is proper time along the particle's worldline $x^a_p = x^a_p(\tau)$.
Specializing to the particle being in a circular geodesic orbit at
areal radius~$r = r_p$, aligning the equator of the coordinate system
($\theta = \tfrac{\pi}{2}$) with the particle orbit, and changing
the variable of integration from proper time~$\tau$ to coordinate
time~$t_\Schw$, this becomes
\begin{equation}
\boxop \Phi
	= - \frac{4 \pi q}{r_p^2}
	  \frac{f_p}{\E_p}
	  \delta(r - r_p)
	  \delta(\theta - \tfrac{\pi}{2})
	  \delta(\varphi - \omega_p t_\Schw)
								      \,\text{,}
\end{equation}
where (and henceforth) the subscript ``$_p$'' denotes evaluation on
the particle's worldline~$r = r_p$.

Now expand $r \Phi$ in spherical harmonics $\{Y_{\ell m}(\theta,\varphi)\}$
(with normalization given by~\eqref{eqn-a_ell-m-defn} below) by defining
the complex scalar fields $\phi_{\ell m} = \phi_{\ell m}(t_\Schw,r)$
such that
\begin{equation}
r \Phi(t_\Schw,r,\theta,\varphi)
	= \sum_{\ell = 0}^\infty \sum_{m = -\ell}^\ell
	  \phi_{\ell m}(t_\Schw,r) \, Y_{\ell m}(\theta,\varphi)
								      \,\text{.}
\end{equation}

Each~$\phi_{\ell m}$ satisfies the inhomogeneous linear wave equation
\begin{equation}
\boxop \phi_{\ell m} + V_\ell(r) \phi_{\ell m}
	= S_{\ell m}(t_\Schw) \delta(r - r_p)
								      \,\text{,}
								\label{eqn-wave}
\end{equation}
where the potential $V_\ell$ and source term $S_{\ell m}$ are given by
\begin{eqnarray}
V_\ell(r)
	& = &	\frac{f(r)}{4}
		\left[ \frac{2M}{r^3} + \frac{\ell(\ell+1)}{r^2} \right]
									\\
S_{\ell m}(t_\Schw)
	& = &	\frac{\pi q f_p^2 a_{\ell m}}{r_p \E_p}
		\exp(-i m \omega_p t_\Schw)
							\label{eqn-S_ell-m-defn}
								      \,\text{,}
\end{eqnarray}
with the (real) coefficients $\{a_{\ell m}\}$ defined by
\begin{widetext}
\begin{subequations}
							\label{eqn-a_ell-m-defn}
\begin{equation}
Y_{\ell m}(\theta{=}\tfrac{\pi}{2}, \varphi) = a_{\ell m} e^{im\varphi}
					       \label{eqn-Y_ell-m-normalization}
								      \,\text{,}
\end{equation}
i.e.,
\begin{equation}
a_{\ell m} = \left\{
	     \begin{array}{ll}
	     \displaystyle
	     (-1)^{(\ell{+}m)/2}
	     \sqrt{\frac{2\ell+1}{4\pi}}
	     \sqrt{\frac{(\ell+m-1)!! \, (\ell-m-1)!!}
			{(\ell+m)!! \, (\ell-m)!!}}
				& \text{if $\ell{-}m$ is even}	\\
	     0			& \text{if $\ell{-}m$ is odd}	
	     \end{array}
	     \right.
								      \,\text{,}
\end{equation}
where the ``double factorial'' function is defined by
\begin{equation}
n!! = \left\{
      \begin{array}{ll}
      n \cdot (n-2)!!	& \text{if $n \ge 2$}		\\
      1			& \text{if $n \le 1$}		
      \end{array}
      \right.
								      \,\text{.}
\end{equation}
\end{subequations}

Each~$\phi_{\ell m}$ can be obtained by numerically solving the
wave equation~\eqref{eqn-wave}.  I discuss the problem domain and
boundary conditions for this equation in
section~\ref{sect-Barack-Ori/problem-domain-and-BCs},
and I discuss the numerical solution in
section~\ref{sect-numerical-methods/numerical-soln-of-wave-eqn}.


\subsection{Computing the Self-Force}
\label{sect-Barack-Ori/self-force}

Assuming that the complex scalar field~$\phi_{\ell m}$ is known for
each~$(\ell,m)$, the contravaraint radial component $F_\self$ of the
$\O(\mu)$~self force may be computed as described by
Barack and Sago~\cite{Barack-Sago-2007}:  For each $\ell \ge 0$, define
\begin{equation}
F_\ell^{(\pm)}(t_\Schw)
	= \sum_{m=-\ell}^\ell
	  Y_{\ell m} \left(
		     \theta{=}\tfrac{\pi}{2}, \varphi{=}\omega_p t_\Schw
		     \right)
	  \left.
	  \frac{\partial \bigl(\phi_{\ell m}/r\bigr)}{\partial r}
	  \right|_{t_\Schw, r{=}r_p^\pm}
								      \,\text{,}
					       \label{eqn-F_ell-full-sum-over-m}
\end{equation}
where $r = r_p^{\pm}$ refers to computing the one-sided derivative as
$r$ approaches the particle worldline either from the outside~($+$) or
the inside~($-$), in both cases on a slice of constant~$t_\Schw$.
For finite-differencing purposes, it's convenient to transform this
derivative into one with respect to $r_*$: since
$\partial r_* \big/ \partial r = 1/f$, we have that
\begin{equation}
\frac{\partial \bigl(\phi_{\ell m}/r\bigr)}{\partial r}
	= \frac{1}{fr} \frac{\partial \phi_{\ell m}}{\partial r_*}
	  -
	  \frac{\phi_{\ell m}}{r^2}
								      \,\text{.}
\end{equation}

Now (following Barack and Lousto~\cite{Barack-Lousto-2005}) observe
that under the transformation $m \to -m$, the wave equation's source
term~$S_{\ell m}$ defined by~\eqref{eqn-S_ell-m-defn} transforms to
its complex conjugate.  Since the wave equation's potential~$V_\ell$
is real and independent of~$m$, this means that the equation's
solution~$\phi_{\ell m}$ also transforms to its complex conjugate.
Thus (using~\eqref{eqn-Y_ell-m-normalization}),
\eqref{eqn-F_ell-full-sum-over-m} simplifies to
\begin{subequations}
							  \label{eqn-F_ell-defn}
\begin{equation}
F_\ell^{(\pm)}(t_\Schw)
	= \sideset{}{'}\sum_{m=0}^\ell
	  a_{\ell m} \exp(im \omega_p t_\Schw)
	  \left.
	  \frac{\partial \bigl(\phi_{\ell m}/r\bigr)}{\partial r}
	  \right|_{t_\Schw, r{=}r_p^\pm}
								      \,\text{,}
\end{equation}
where for any quantities $X_{\ell m}$, we define the notation
\begin{equation}
\sideset{}{'}\sum_{m=0}^\ell X_{\ell m}
	\eqdef	X_{\ell 0} + 2 \sum_{m=1}^\ell \Realpart[X_{\ell m}]
								      \,\text{.}
\end{equation}
\end{subequations}
\end{widetext}

Following Barack and Ori~\cite{Barack-Ori-2002}, the contravariant
radial component of the self-force at any point on the particle's
worldline is then given by
\begin{subequations}
							   \label{eqn-F_self-pm}
\begin{equation}
F_\self^{(\pm)}(t_\Schw) = \sum_{\ell=0}^\infty F^{(\pm)}_{\ell,\reg}(t_\Schw)
								      \,\text{,}
						   \label{eqn-F_self-pm-sum-ell}
\end{equation}
where the regularized self-force modes $F^{(\pm)}_{\ell,\reg}$ are
given by
\begin{equation}
F^{(\pm)}_{\ell,\reg}(t_\Schw)
	= F_\ell^{(\pm)}(t_\Schw) \mp (\ell+\thalf) A(r_p) - B(r_p)
								      \,\text{,}
\end{equation}
\end{subequations}
where (for a particle in a circular geodesic orbit in Schwarzschild
spacetime) the regularization coefficients $A(r)$ and $B(r)$ are
given by
\begin{subequations}
					    \label{eqn-AB-regularization-coeffs}
\begin{eqnarray}
A(r)	& = &	\frac{q^2}{r^2} \frac{\E}{f\mathcal{V}}
									\\
B(r)	& = &	\frac{q^2}{r^2}
		\frac{\E^2 [\hat{E}(w) - 2 \hat{K}(w)]}
		     {\pi f \mathcal{V}^{3/2}}
								      \,\text{,}
\end{eqnarray}
\end{subequations}
where $\mathcal{V}$ and $w$ are given by
\begin{eqnarray}
\mathcal{V}	& = &	1 + \L^2/r^2				\\
w		& = &	\frac{\L^2}{\L^2 + r^2}
								      \,\text{,}
\end{eqnarray}
and $\hat{K}(w)$ and $\hat{E}(w)$ are the complete elliptic
integrals of the first and second kinds respectively,
\begin{subequations}
					\label{eqn-Khat-Ehat-elliptic-integrals}
\begin{eqnarray}
\hat{K}(w)	& = &	\int_0^{\pi/2} \frac{1}{\sqrt{1 - w \sin^2 x}} \, dx
									\\
\hat{E}(w)	& = &	\int_0^{\pi/2} \sqrt{1 - w \sin^2 x} \, dx
								      \,\text{.}
\end{eqnarray}
\end{subequations}

Barack and Ori~\cite{Barack-Ori-2002} have shown that
$F^{(+)}_{\ell,\reg} = F^{(-)}_{\ell,\reg}$ and hence that
$F_\self^{(+)} = F_\self^{(-)}$.  In view of this the $^{(+)}$
and $^{(-)}$ superscripts may be dropped, and we may
rewrite~\eqref{eqn-F_self-pm-sum-ell} as
\begin{equation}
F_\self = \sum_{\ell = 0}^\infty F_{\ell,\reg}
						      \label{eqn-F_self-sum-ell}
\end{equation}
without ambiguity.  However, for numerical purposes it's
still very useful to compute both expressions $F^{(+)}_{\ell,\reg}$
and~$F^{(-)}_{\ell,\reg}$; I discuss this in
section~\ref{sect-numerical-methods/internal-error-estimates}.


\subsection{Problem Domain and Boundary Conditions}
\label{sect-Barack-Ori/problem-domain-and-BCs}

The wave equation~\eqref{eqn-wave} is naturally posed on an infinitely
large domain with boundary conditions at infinity appropriate for an
isolated system in an asymptotically flat spacetime.  However, for
numerical purposes it's convenient to instead follow an approach
suggested by Barack and Lousto~\cite{Barack-Lousto-2005},
solving~\eqref{eqn-wave} on a large but finite domain using arbitrary
initial data and/or boundary conditions.  These introduce a burst
of spurious ``radiation'' dynamics into the solution~$\phi_{\ell m}$,
but fortunately this spurious radiation dies out quite quickly as
one moves away from the initial slice(s) and/or the problem-domain
boundaries.
\footnote{
	 I have seen no evidence of the Jost ``persistent junk''
	 solutions discussed by~\cite{Field-Hesthaven-Lau-2010}.
	 }
{}  The self-force is defined along the particle's worldline, and
its value at a given event~$\Q$ on that worldline depends only on
$\phi_{\ell m}$ and $\del \phi_{\ell m}$ at~$\Q$.  The effect of
the spurious radiation can thus be made negligible by choosing a
sufficiently large numerical problem domain whose initial slice
and/or boundaries are sufficiently distant from~$\Q$.


\subsection{The Tail Sum}
\label{sect-Barack-Ori/tail-force}

The definition~\eqref{eqn-F_self-sum-ell} of the self-force $F^{(\pm)}$
is written in terms of an infinite sum $\sum_{\ell = 0}^\infty$ of
regularized self-force modes $F_{\ell,\reg}$.  For numerical purposes
a finite expression is needed.  Following
Barack and Sago~\cite[section~III.E]{Barack-Sago-2007}, partition the
infinite sum~\eqref{eqn-F_self-sum-ell} into a finite ``numerical force''
sum of the modes with $\ell \le K$ and an infinite ``tail force'' sum
of the modes with $\ell \ge K' \equiv K{+}1$, where $K \sim 30$ is a
numerical parameter:
\begin{subequations}
				       \label{eqn-self-force/numerical+tail-force}
\begin{eqnarray}
F_\self
	& = &	F_{\self,\num} + F_{\self,\tail}
					   \label{eqn-self-force/numerical+tail}
									\\
F_{\self,\num}
	& = &	\sum_{\ell=0}^K F_{\ell,\reg}
					  \label{eqn-self-force/numerical-force}
									\\
F_{\self,\tail}
	& = &	\sum_{\ell=K'}^\infty F_{\ell,\reg}
					       \label{eqn-self-force/tail-force}
								      \,\text{.}
\end{eqnarray}
\end{subequations}

Once the regularized self-force modes $F_{\ell,\reg}$ are known for
$0 \le \ell \le K$, the numerical force~$F_{\self,\num}$ is easy to
compute from the definition~\eqref{eqn-self-force/numerical-force}.
The tail force~$F_{\self,\tail}$ can be estimated using the known
large-$\ell$ series
expansion~\cite[equation~(12)]{Detweiler-Messaritaki-Whiting-2003}
\begin{equation}
F_{\ell,\reg} = \sum_{\substack{\text{$p$ even} \\ {p \ge 2}}} c_p f_p(\ell)
							 \label{eqn-tail-series}
								      \,\text{,}
\end{equation}
where the $\{c_p\}$ are coefficients not depending on~$\ell$, and
the basis functions $f_p(\ell) = \O(\ell^{-p})$ are given by
\begin{widetext}
\begin{subequations}
						      \label{eqn-tail-force-basis}
\begin{eqnarray}
f_2(\ell)
	& = &	\frac{1}
		     {(\ell-\frac{1}{2})(\ell+\frac{3}{2})}
									\\
f_4(\ell)
	& = &	\frac{1}
		     {(\ell-\frac{3}{2})(\ell-\frac{1}{2})
		      (\ell+\frac{3}{2})(\ell+\frac{5}{2})}
									\\
f_6(\ell)
	& = &	\frac{1}
		     {(\ell-\frac{5}{2})(\ell-\frac{3}{2})(\ell-\frac{1}{2})
		      (\ell+\frac{3}{2})(\ell+\frac{5}{2})(\ell+\frac{7}{2})}
									\\
f_8(\ell)
	& = &	\frac{1}
  {(\ell-\frac{7}{2})(\ell-\frac{5}{2})(\ell-\frac{3}{2})(\ell-\frac{1}{2})
   (\ell+\frac{3}{2})(\ell+\frac{5}{2})(\ell+\frac{7}{2})(\ell+\frac{9}{2})}
									\\
	& \cdots &
		\phantom{
		  \frac{1}
		       {(\ell-\frac{1}{2})(\ell+\frac{3}{2})}
			}
								\nonumber
\end{eqnarray}
\end{subequations}
Typically only a few terms in this series are needed to give an
excellent approximation to $F_{\ell,\reg}$.

For a particle in a circular geodesic orbit in Schwarzschild spacetime,
Detweiler, Messaritaki, and Whiting~\cite{Detweiler-Messaritaki-Whiting-2003}
have shown that the coefficient~$c_2$ is given by
\begin{eqnarray}
c_2 = - \frac{1}{4}
	\cdot 2 \sqrt{2}
	\sqrt{\frac{2 r_p^2 (r_p - 2M)}{r_p - 3M}}
	&   &
		\Biggl[
		{}
		- \frac{M(r_p - 2M)}{2r_p^4(r_p - 3M)} G_{-1/2}
		- \frac{(r_p - M)(r_p - 4M)}{8 r_p^4 (r_p - 2M)} G_{1/2}
								\nonumber\\
	&   &	\phantom{\Biggl[}
		{}
		+ \frac{(r_p - 3M)(5 r_p^2 - 7 r_p M - 14 M^2)}
		       {16 r_p^4 (r_p - 2M)^2}
		  G_{3/2}
		\phantom{\Biggr]}
								\nonumber\\
	&   &	\phantom{\Biggl[}
		{}
		- \frac{3 (r_p - 3M)^2 (r_P + M)}{16 r_p^4 (r_p - 2M)^2} G_{5/2}
		\Biggr]
						  \label{eqn-c_2-tail-fit-coeff}
\end{eqnarray}
\end{widetext}
where the leading factor of~$-1/4$ converts from the normalization
used by Detweiler, Messaritaki, and Whiting to that used here, and
where $G_p$ (a special case of a Gauss hypergeometric function)
is given by
\begin{equation}
G_p = \frac{2}{\pi}
      \int_0^{\pi/2} (1 - \alpha \sin^2 x)^{-p} \, dx
							    \label{eqn-G_p-defn}
								      \,\text{,}
\end{equation}
with $\alpha = M / (r_p - 2M)$.  $G_{\pm 1/2}$ can also be written
in terms of the complete elliptic
integrals~\eqref{eqn-Khat-Ehat-elliptic-integrals},
\begin{subequations}
			  \label{eqn-G_{pm-1/2}-as-Khat-Ehat-elliptic-integrals}
\begin{eqnarray}
G_{-1/2}	& = &	\frac{2}{\pi} \hat{E}(\alpha)			\\
G_{1/2}		& = &	\frac{2}{\pi} \hat{K}(\alpha)
								      \,\text{.}
\end{eqnarray}
\end{subequations}

The $c_4$ and higher coefficients aren't known analytically, but they
can be estimated numerically by least-squares fitting the tail-series
expansion~\eqref{eqn-tail-series} to some suitable subset of the
numerically-computed $F_{\ell,\reg}$ values.  I discuss the numerical
computation of this ``tail fit'' in
section~\ref{sect-numerical-methods/tail-fit}.

Once the $\{c_p\}$~coefficients are known, the
tail force~\eqref{eqn-self-force/tail-force} is then given by
\begin{equation}
F_{\self,\tail}
	= \sum_{\ell=K'}^\infty F_{\ell,\reg}
	= \sum_{\substack{\text{$p$ even} \\ {p \ge 2}}} c_p \Gamma_p
							    \label{eqn-tail-force}
								      \,\text{,}
\end{equation}
where
\begin{equation}
\Gamma_p = \sum_{\ell=K'}^\infty f_p(\ell)
								      \,\text{.}
						 \label{eqn-tail-force-Gamma-defn}
\end{equation}
Using the Maple symbolic algebra system~(\cite{Char-etal-1983:Maple-design},
\url{http://www.maplesoft.com/}, version~11) to evaluate the
sums~\eqref{eqn-tail-force-Gamma-defn},
\footnote{
	 These sums can also be evaluated by hand by first
	 using partial fractions, after which each sum
	 telescopes, then finally undoing the partial
	 fractions to further simplify the result.  I have
	 explicitly verified~\eqref{eqn-Gamma2}, \eqref{eqn-Gamma4},
	 and~\eqref{eqn-Gamma6} in this way.
	 }
{} I find that the first few~$\Gamma_p$ are given by
\begin{widetext}
\begin{subequations}
						      \label{eqn-tail-force-Gamma}
\begin{eqnarray}
\Gamma_2
	& = &	\frac{K'}{(K'-\frac{1}{2}) (K'+\frac{1}{2})}
							      \label{eqn-Gamma2}
									\\
\Gamma_4
	& = &	\frac{K'}{3 (K'-\frac{3}{2}) (K'-\frac{1}{2})
			    (K'+\frac{1}{2}) (K'+\frac{3}{2})}
							      \label{eqn-Gamma4}
									\\
\Gamma_6
	& = &	\frac{K'}
		     {5 (K'-\frac{5}{2}) (K'-\frac{3}{2}) (K'-\frac{1}{2})
			(K'+\frac{1}{2}) (K'+\frac{3}{2}) (K'+\frac{5}{2})}
							      \label{eqn-Gamma6}
									\\
\Gamma_8
	& = &	\frac{K'}
  {7 (K'-\frac{7}{2}) (K'-\frac{5}{2}) (K'-\frac{3}{2}) (K'-\frac{1}{2})
     (K'+\frac{1}{2}) (K'+\frac{3}{2}) (K'+\frac{5}{2}) (K'+\frac{7}{2})}
								      \,\text{.}
\end{eqnarray}
\end{subequations}
\end{widetext}


\section{Numerical Computation of the Self-Force}
\label{sect-numerical-methods}

In this section I describe the numerical methods I use for
high-accuracy self-force calculations.


\subsection{Numerical Solution of the Wave Equation~(\protect\ref{eqn-wave})}
\label{sect-numerical-methods/numerical-soln-of-wave-eqn}


\subsubsection{General Numerical Scheme}
\label{sect-numerical-methods/numerical-soln-of-wave-eqn/general-num-scheme}

Near the particle worldline the complex scalar field $\phi_{\ell m}$
has $\O(1)$~amplitude and rapidly oscillating phase in both space
and time, but the field amplitude decreases quickly with increasing
distance from the particle worldline.  This high dynamic range suggests
the use of a mesh-refinement method to resolve the fast oscillations
without the computational cost of of maintaining this high resolution
everywhere in the numerical domain.  The numerical method also
needs to accommodate the non-differentiability of $\phi_{\ell m}$
across the particle worldline.

\begin{figure}[bp]
\begin{center}
\includegraphics[scale=1.10]{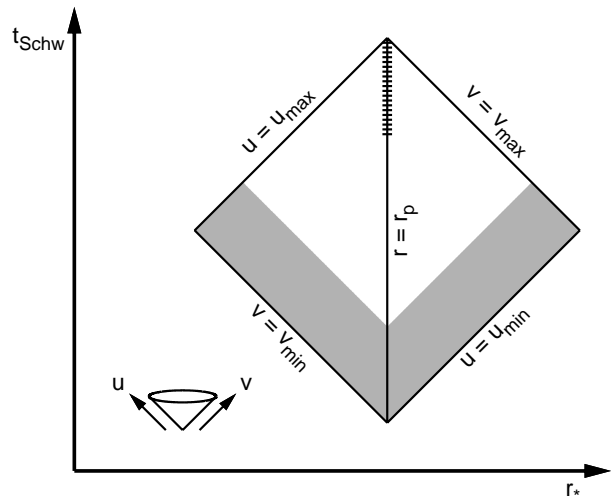}
\end{center}
\vspace{-5mm}
\caption[Numerical Problem Domain]
	{
	This figure shows the overall problem domain,
	and the $(u,v)$ and $(t_\Schw,r_*)$ coordinates.
	The vertical line marks the particle worldline.
	Mesh refinement is inhibited in the V-shaped shaded
	region $100M$ wide bordering the ``southeast'' and
	``southwest'' grid faces.  The self-force is measured
	along the region of the particle worldline marked by
	horizontal hatching.
	}
\label{fig-problem-domain}
\end{figure}

To avoid the numerical complications of explicit boundary conditions,
I follow Barack and Lousto~\cite{Barack-Lousto-2005} and use a characteristic
(double-null) numerical evolution scheme, with a ``diamond-shaped''
problem domain which is a square in the characteristic variables~$u$
and $v$, $(u,v) \in [u_{\min},u_{\max}] \times [v_{\min},v_{\max}]$.
With this domain the (arbitrary) initial data $\phi_{\ell m} = 0$
is applied on the ``southwest'' and ``southeast`` grid faces
$v = v_{\min}$ and $u = u_{\min}$ respectively; I place the domain
such that the particle worldline $r = r_p$ symmetrically bisects
the domain.  Figure~\ref{fig-problem-domain} illustrates the
problem domain and particle worldline.  This type of problem setup
has been used successfully for a number of other self-force calculations,
including (for example) those of~\cite{Haas-2007,Barack-Sago-2010}.

To numerically solve the wave equation~\eqref{eqn-wave} on this
domain, I use a characteristic adaptive mesh refinement (AMR)
numerical scheme with 4th~order global finite differencing
accuracy.  I have described this scheme in detail
elsewhere~\cite{Thornburg-2009:characteristic-AMR}.  Briefly,
the underlying (unigrid) finite differencing is a standard
double-null diamond-integral scheme with square grid cells in $(v,u)$
(\cite{Gomez-Winicour-1992:sssf-2+2-evolution-and-asymptotics,
Gomez-Winicour-Isaacson-1992:sssf-2+2-numerical-methods,
Gundlach-Price-Pullin-1994a,Burko-Ori-1997,Lousto-Price-1997,
Lousto-2005,Winicour-2009:living-review}),
extended to provide globally 4th~order finite differencing
accuracy in a manner similar to that of~\cite{Lousto-2005,Haas-2007}.
The AMR algorithm it is very similar to the standard Cauchy Berger-Oliger
AMR algorithm (\cite{Berger-1984}; see also \cite{Berger-1982,Berger86,
Berger-1989,Choptuik89,Choptuik-in-d'Inverno:self-similarity-and-AMR}),
slightly modified as suggested by
Hamad\'{e} and Stewart~\cite{Hamade-Stewart-1996}
to accommodate the characteristic evolution.  The AMR algorithm
treats treats $v$~as a ``time'' coordinate and $u$~as a ``space''
coordinate: the evolution integrates $v = \text{constant}$ slices
successively in the direction of increasing~$v$, with each slice
completely integrated (in the direction of increasing~$u$) before
the integration of the next slice begins.

The AMR algorithm begins with a relatively coarse ``base'' grid
which covers the entire problem domain; during the evolution the
algorithm dynamically (adaptively) constructs a hierarchy of finer
``child'' grids, each a factor of~$2$ finer than, and spatially
nested inside, its ``parent'' grid.  The fine grids typically only
cover small subsets of the problem domain.  

The AMR algorithm is controlled (the "adaptive" part of AMR) by
comparing an estimate~$\Lambda$ of the numerical solution's local
truncation error (LTE)
\footnote{
	 The LTE is a measure of the \emph{local} accuracy
	 with which the finite difference equations approximate
	 the underlying PDE (here the wave equation~\eqref{eqn-wave}).
	 More precisely, the LTE is a pointwise norm of the
	 discrepancy that would result if the exact solution
	 of the PDE were substituted into the finite difference
	 equations at a grid
	 point~\cite{Kreiss73,Choptuik-1991:FD-consistency,
Richtmyer-Morton-2nd-edition,LeVeque-2007:finite-diff-methods}.
	 }
{} with a specified tolerance $\AMRtolerance$.  If (after median
smoothing~\cite[section~4.2]{Thornburg-2009:characteristic-AMR})
$\Lambda > \AMRtolerance$ then the algorithm adds another level of
mesh-refinement to better resolve the solution.

As well as AMR, the numerical scheme also uses fixed mesh refinement
(FMR): following~\cite{Choptuik-in-d'Inverno:self-similarity-and-AMR},
my AMR code has options to record the placement and grid spacing of
each refinement level generated by the AMR algorithm.  This can then
be ``played back'' with each grid refined by a chosen small-integer
factor $N_\fmr$.  FMR is useful both for convergence tests, and in
some cases for circumventing floating-point roundoff limits on my
AMR scheme (these are discussed in
section~\ref{sect-numerical-methods/numerical-soln-of-wave-eqn/extended-fp-precision}).

Because of the characteristic evolution scheme, the local finite
differencing must actually be 6th~order accurate in order to achieve
a 4th~order global accuracy
(see~\cite[section 3.1]{Thornburg-2009:characteristic-AMR} and
references therein).  Similarly, the global accuracy generally
scales as $\AMRtolerance^{2/3}$, or $\ErrorToleranceEffective^{2/3}$ 
if FMR is used, where the ``effective error tolerance'' is
$\ErrorToleranceEffective \eqdef \AMRtolerance / N_\fmr^6$
(cf.~discussion in section~\ref{sect-results/tail-fits},
particular figure~\ref{fig-tail-fit-Chi2-residuals-errors}).


\subsubsection{Extended Floating-Point Precision}
\label{sect-numerical-methods/numerical-soln-of-wave-eqn/extended-fp-precision}

Floating-point numbers are only represented and computed with finite
accuracy; typically each floating-point operation introduces a small
roundoff error of fractional size ${\ltsim}\, \FPepsilon$, where
$\FPepsilon = 2^{-52} \approx \SciNum{2.2}{-16}$ for IEEE-standard
double-precision floating-point arithmetic.
\footnote{
	 More precisely, $\FPepsilon$, usually known as the
	 ``machine epsilon'', is defined as the smallest positive
	 floating-point number such that $1 \oplus \FPepsilon \ne 1$,
	 where $\oplus$ is the floating-point addition operator.
	 This is discussed in detail by, for example,
         \cite[chapter~2]{FMM77}; \cite[chapter~2]{Kahaner-Moler-Nash-1989};
	 and \cite{Goldberg91} and references therein.
	 }

There are (at least) two different parts of my numerical scheme for
solving the wave equation~\eqref{eqn-wave} which may be limited in
accuracy by floating-point roundoff effects:
\begin{enumerate}
\item[\FPlimitPerGridPoint]
	The first and most obvious way in which floating-point
	roundoff effects limits the achievable accuracy of the
	numerical scheme is the finite-difference computation
	of $\phi_{\ell m}$ at each successive grid point.
	This computation
(described in detail in~\cite[appendix~A.2]{Thornburg-2009:characteristic-AMR})
	involves ${\sim}\, 50$~floating-point operations.  In the
	absence of fortuituous error cancellations, this computation
	contributes a relative error of $\sigma \FPepsilon$ at each
	grid point, where $\sigma \gtsim 1$ reflects the error-propagation
	properties of the computation (which I have not analyzed in detail).

\item[\FPlimitLTEEstimate]
	The second way in which floating-point roundoff may limit
	the achievable precision of my numerical scheme is via the
	AMR algorithm:  My code estimates the LTE by comparing the
	standard numerical computation of $\phi$ at a grid point
	with an alternate lower-resolution computation which spans
	the most recent 2~grid points in~$v$ and~$u$ with a single
	finite differencing
	step~\cite[equation~(6)]{Thornburg-2009:characteristic-AMR}.
	If the difference between $\phi$ computed in these two ways
	isn't well resolved by the floating-point arithmetic, the
	LTE estimate will be unreliable.
\footnote{
	 If $\Lambda$ is unreliable, then (even after the smoothing)
	 we might well have $\Lambda > \AMRtolerance$ somewhere on
	 each new slice, no matter how small the grid spacing.  This
	 would cause the AMR algorithm to effectively infinite-loop,
	 continually adding further refinement levels until it runs
	 out of memory.  Although a limit on the maximum refinement
	 level could prevent this, the algorithm would still be
	 refining inappropriately, causing the computation to be
	 very inefficient.
	 }
{}	In practice, taking into account the normalization factors
	in the actual LTE estimate, I ensure reliable operation of
	the AMR algorithm by limiting it to an LTE-estimate tolerance
	$\AMRtolerance \gtsim \FPepsilon$.
\end{enumerate}

One way to circumvent the AMR LTE-estimate limit~\FPlimitLTEEstimate{}
is (following~\cite{Choptuik-in-d'Inverno:self-similarity-and-AMR})
to record the placement and grid spacing of each refinement level
generated by the AMR algorithm, then ``play back'' this with
each grid refined by a chosen small-integer factor $N_\fmr$.  This
``fixed mesh refinement'' (FMR) reduces the global finite-difference
truncation error (the cumulative effects of the LTE in all the grid
cells in the entire numerical integration) by very close to a factor
of~$N_\fmr^4$~\cite[figure~7]{Thornburg-2009:characteristic-AMR},
at the cost of an increase in the code's running time by a factor
of~$N_\fmr^2$.  However, the per-grid point rounding error
limit~\FPlimitPerGridPoint{} cannot be circumvented in this way.
(In fact, FMR may \emph{worsen} the overall floating-point roundoff
errors in the self-force by a factor of~$N_\fmr^2$ or more due to the
larger number of individually-smaller grid cells in the integration.)

Due to the AMR LTE-estimate limit~\FPlimitLTEEstimate{}, I restrict
the AMR algorithm to a tolerance $\AMRtolerance \gtsim 10^{-16}$ when
using standard IEEE double-precision floating-point arithmetic.  FMR
can improve this considerably, but beyond $N_\fmr \approx 6$, the
per-grid-point rounding error limit~\FPlimitPerGridPoint{} becomes
increasingly severe, and my error estimates for the individual
$F_{\ell,\reg}$, the tail fit, and the overall self-force all become
less reliable.

To further investigate the effects of floating-point rounding errors
in the numerical solution of the wave equation~\eqref{eqn-wave}, I
extended-precision floating-point arithmetic.  In particular, on
Intel~x86 and compatible processors my AMR code for solving the
wave equation~\eqref{eqn-wave} can optionally use IEEE ``double-extended''
floating-point arithmetic (typically specified in C/\Cplusplus{} as
``\texttt{long double}'').  This provides a relative accuracy of
$\FPepsilon = 2^{-63} \approx \SciNum{1.1}{-19}$, a factor of
$2^{11} = 2048$~times more accurate than IEEE double precision.
This lowers the AMR LTE-estimate limit~\FPlimitLTEEstimate{} to
$\AMRtolerance \gtsim  10^{-19}$, with only a modest performance
penalty compared to standard IEEE double precision (at the same
accuracy setting my code is about a factor of~$2$~slower in
long-double than in double precision).

Note that even when using extended-precision arithmetic in this way,
once the gradients
$\left.
 \partial (\phi_{\ell m}/r) \big/ \partial r
 \right|_{t_\Schw, r{=}r_p^\pm}$
are known along the particle worldline, the remainder of the self-force
computation is considerably less sensitive to the floating-point
arithmetic precision.  I thus use standard IEEE double precision
for computing each regularized self-force $F_{\ell,\reg}^{\pm}$,
the numerical force~\eqref{eqn-self-force/numerical-force},
the tail fit and tail force~\eqref{eqn-tail-force}, and the error
estimates for these quantities.
\footnote{
	 To (slightly) reduce floating-point roundoff errors,
	 I use Kahan summation~(\cite{Kahan-1965:accurate-summation};
	 \cite[theorem~8]{Goldberg91}) when evaluating the
	 sums~\eqref{eqn-F_ell-defn}, \eqref{eqn-F_self-pm-sum-ell}
	 and~\eqref{eqn-tail-force}.
	 }


\subsection{Parallel Execution}
\label{sect-numerical-methods/parallel}

Even with AMR, self-force computations are still very expensive,
so it's useful to parallelize them as much as possible.  Fortunately,
the self-force problem is trivially parallelizable by distributing
the solution of the wave equation~\eqref{eqn-wave} to different
processors for different $(\ell,m)$.  Because no communication
is needed between the computations for different $(\ell,m)$, this
requires very little communications bandwidth, and overall performance
scales almost linearly with the number of processors used.
\footnote{
	 In the parallel-computing community, this type of
	 problem is known as ``embarrassingly parallel'',
	 in the sense that it's such an easy test case for
	 parallel hardware that one should be embarrassed
	 to report parallel-speedup results for it.
	 }

For the results presented here, I used between 10 and 15~processors
of a local workstation cluster, with a shared NFS file system to collect
the results from each processor's computations.  Each processor was
either a 2.5~GHz, 2.8~GHz, or 3.2~GHz Pentium~4.


\subsection{Regularization Coefficients}
\label{sect-numerical-methods/reg-coeffs}

I compute the regularization coefficients~$A(r)$ and $B(r)$ from
the definitions~\eqref{eqn-AB-regularization-coeffs}, evaluating
the $\hat{K}$ and $\hat{E}$ complete elliptic
integrals~\eqref{eqn-Khat-Ehat-elliptic-integrals} using the
\subroutine{ellpk} and \subroutine{ellpe} subroutines from the
\SubroutineLibrary{Cephes} library (\cite{Moshier-1989},
\url{http://www.netlib.org/cephes}, release~2.2 dated July~1992).


\subsection{The Tail Fit}
\label{sect-numerical-methods/tail-fit}

I consider two cases for the tail fit:
\begin{itemize}
\item	For the most accurate computation possible (assuming
	the particle to be in a circular geodesic orbit),
	I compute the tail-fit coefficient~$c_2$ from the
	expression~\eqref{eqn-c_2-tail-fit-coeff}, evaluating
	each $G_p$ via direct numerical integration of the
	definition~\eqref{eqn-G_p-defn}, using the \subroutine{dqags}
	subroutine (revision date 1983 May~18) from the
	\SubroutineLibrary{Quadpack} library
	(\cite{Piessens-etal-1983:Quadpack-book},
	\url{http://www.netlib.org/quadpack}).
\footnote{
	 I have also explicitly verified that the
	 identities~\eqref{eqn-G_{pm-1/2}-as-Khat-Ehat-elliptic-integrals}
	 hold to very high accuracy (a few parts in $10^{16}$)
	 for my numerical implementation.
	 }

	As noted in section~\ref{sect-Barack-Ori/tail-force}, the
	$c_4$~and higher tail-fit coefficients can be estimated
	numerically by least-squares fitting the series
	expansion~\eqref{eqn-tail-series} to some suitable
	subset of the numerically-computed $F_{\ell,\reg}$ values.
	For the accuracies obtained in this paper, it suffices to
	keep only terms up to and including the $\O(\ell^{-6})$ term,
	so the tail fit only includes the coefficients $\{c_4,c_6\}$.

\item	Alternatively, to simulate the accuracy to be expected
	for a particle in a generic non-circular orbit (where the
	$c_2$~coefficient isn't known analytically for the form
	of the mode-sum regularization used here),
\footnote{
	 Haas and Poisson~\cite{Haas-Poisson-2006} have
	 computed the equivalent of the $c_2$~coefficient
	 for a different form of mode-sum regularization,
	 and Haas~\cite{Haas-2007} has used this successfully
	 in a numerical computation of the self-force on a
	 scalar particle in a generic (non-circular) orbit
	 in Schwarzschild spacetime.
	 }
{}	I also consider the case where $c_2$ is included in the
	tail fit, i.e., where the coefficients $\{c_2,c_4,c_6\}$
	are fitted simultaneously.
\end{itemize}

Whichever set of coefficients are fitted, computing the
tail fit numerically requires some care, because the basis
functions~$\{f_p\}$ defined by~\eqref{eqn-tail-force-basis} are
nearly degenerate (linearly dependent), causing the tail fit to
be quite ill-conditioned.  That is, there are linear combinations
of the basis functions $\sum_p b_p f_p$ where the linear-combination
coefficients $\{b_p\}$ have unit 2-norm (call these ``unit-coefficient-norm''
linear combinations), yet the linear combination $\sum_p b_p f_p$
is very small relative to the largest of the $\{f_p\}$.  The fitted
coefficients $\{c_p\}$ are relatively uncertain in the direction
of any such $\{b_p\}$, which introduces additional uncertainty into
the tail force~$F_{\self,\tail}$ computed via~\eqref{eqn-tail-force}.

\begin{figure}[bp]
\begin{center}
\hspace*{-1mm}
\includegraphics[scale=1.10]{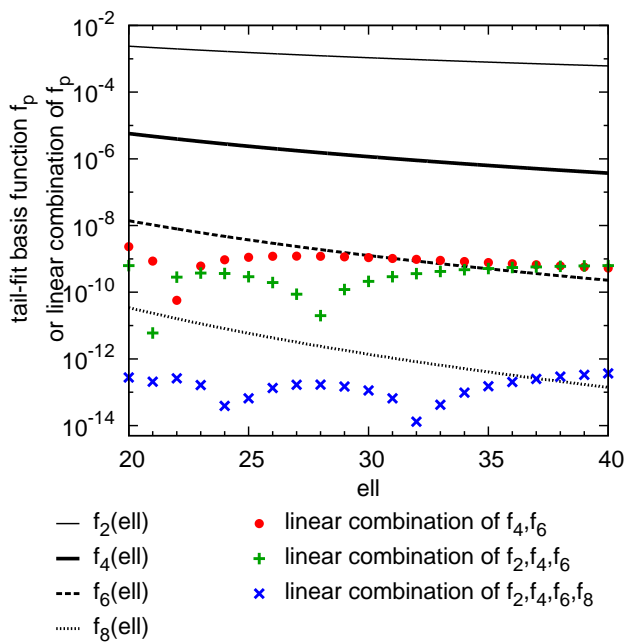}
\end{center}
\vspace{-5mm}
\caption[Near-Degeneracy of the Tail-Fit Basis Functions $\{f_p\}$]
	{
	[Color online]
	This figure shows the tail-fit basis functions $f_2$, $f_4$,
	$f_6$, and $f_8$, together with 3~very small unit-coefficient-norm
	linear combinations of the basis functions, $\sum_p b_p f_p$.
	}
\label{fig-tail-fit-basis-degeneracy}
\end{figure}

Figure~\ref{fig-tail-fit-basis-degeneracy} illustrates the near-degeneracy
of the $\{f_p\}$, showing very small unit-coefficient-norm
linear combinations of various subsets of the $\{f_p\}$, and
table~\ref{tab-tail-fit-basis-degeneracy} gives the corresponding
condition numbers~$\kappa$.
\footnote{
	 These very-small linear combinations can be determined from
	 a singular value decomposition (SVD) of the least-squares
	 fit's design matrix (\cite{Lawson-Hanson-1974:least-squares};
\cite[chapter~9]{FMM77}; \cite[section~6.8]{Kahaner-Moler-Nash-1989};
\cite{Hammarling-1985:SVD-in-multivariate-statistics};
\cite[section~15.4]{Numerical-Recipes-2nd-edition};
\cite{Bjoerck-1996:algorithms-for-least-squares-problems}).
	 For present purposes the condition number~$\kappa$
	 can be interpreted as the ratio of the largest 2-norm
	 of any basis function to the smallest 2-norm of any
	 unit-coefficient-norm linear combination $\sum_p b_p f_p$
	 of the basis functions.  Thus $1 \le \kappa \le \infty$,
	 with $\kappa = 1$ describing an orthonormal basis set,
	 $\kappa \gg 1$ describing a nearly degenerate basis set,
	 and $\kappa = \infty$ describing a perfectly degenerate
	 (linearly dependent) basis set.  Small errors in the
	 input data and/or computation of the fit -- including
	 in particular floating-point roundoff errors -- are
	 amplified by a factor proportional to $\kappa$ in the
	 outputs of the fit (the fitted coefficients $\{c_p\}$),
	 and thus also in the tail force~$F_{\self,\tail}$ computed
	 via~\eqref{eqn-tail-force}.
	 }
{}  The condition numbers are primarily determined by how many
coefficients~$\{c_k\}$ are fit simultaneously: fitting 2, 3, or
4~coefficients simultaneously gives a condition number of
of~$\kappa \sim 10^3$, $10^6$, or $10^9$~respectively.

\begin{table}[bp]
\begin{center}
\begin{ruledtabular}
\begin{tabular}{lcd}
Coefficients		& \multicolumn{2}{c}{Condition Number $\kappa$}	\\
\cline{2-3}								\\[-2ex]
Being Fitted		& Basis is $\{f_p\}$	&
				\multicolumn{1}{c}{Basis is $\{\normalize{f}_p\}$}
									\\
\hline									\\[-2ex]
$\{c_4,c_6\}$		& $\SciNum{2.6}{3}$	&  11			\\
$\{c_4,c_6,c_8\}$	& $\SciNum{5.9}{6}$	& 100			\\
\hline									\\[-2ex]
$\{c_2,c_4\}$		& $\SciNum{2.2}{3}$	&   8.3			\\
$\{c_2,c_4,c_6\}$	& $\SciNum{4.4}{6}$	&  66			\\
$\{c_2,c_4,c_6,c_8\}$	& $\SciNum{8.0}{9}$	& 460			\\
\end{tabular}
\end{ruledtabular}
\end{center}
\caption[Condition Numbers of the Tail-Fit Basis]
	{
	This table shows the condition number~$\kappa$ of the
	tail fit (more precisely, of the fit's design matrix
	if all the regularized self-force modes $F_{\ell,\reg}$
	are taken to have unit uncertanties) when various sets
	of tail-fit coefficients $\{c_p\}$ are fitted and
	either the $\{f_p\}$ or $\{\normalize{f}_p\}$ basis functions
	are used in the fit.  The $F_{\ell,\reg}$ are assumed
	to be given for $20 \le \ell \le 30$, $\ell=35$, and
	$\ell=40$, as is the case for the numerical results
	presented in section~\ref{sect-results}.  However, the
	condition numbers depend only relatively weakly on the
	precise set of $\ell$ used and the relative uncertainties
	of the different $F_{\ell,\reg}$.
	}
\label{tab-tail-fit-basis-degeneracy}
\end{table}

Because of this ill-conditioning, it's much better (yields more
accurate results) to perform the tail fit using a $\mathsf{QR}$ or
singular value decomposition, rather than via the normal equations
(\cite[section~2.2]{Bjoerck-1996:algorithms-for-least-squares-problems}).
(If the normal equations were used, the effective condition number
would be roughly the square of that given here, thus greatly
increasing the effects of floating-point roundoff errors on the
results.)

Much of this ill-conditioning is due to the widely differing
magnitudes of the different basis functions (this can be seen in
figure~\ref{fig-tail-fit-basis-degeneracy}), and can be greatly
alleviated by simply renormalizing the basis functions to have
similar magnitudes over the range of $\ell$ used in the tail fit.
To this end, I define
\begin{equation}
\normalize{f}_p(\ell) = \frac{f_p(\ell)}{f_p(\normalize{\ell})}
								      \,\text{,}
					   \label{eqn-tail-force-basis-normalized}
\end{equation}
where the parameter $\normalize{\ell}$ (taken here to be $20$) is the
$\ell$ at which all the normalized basis functions will now have
unit magnitude.  Table~\ref{tab-tail-fit-basis-degeneracy} also
gives the condition number for fits using various subsets of the
normalized basis functions~$\{\normalize{f}_p\}$.  The normalized basis
sets have much smaller condition numbers, and correspondingly lead
to significantly smaller floating-point roundoff effects in the
tail fits.

I compute the fitted coefficients $\{c_p\}$ and their covariance
matrix using the \subroutine{gsl\_multifit\_wlinear\_svd} subroutine
from the \SubroutineLibrary{GNU Scientific Library}
(\cite{Galassi-etal-2009:GSL-manual}, version~1.12),
using the normalized basis functions $\{\normalize{f}_p(\ell)\}$
defined by~\eqref{eqn-tail-force-basis-normalized}.
\footnote{
	 \subroutine{gsl\_multifit\_wlinear\_svd}
	 actually does its own scaling internally, similar to
	 my normalization~\eqref{eqn-tail-force-basis-normalized}.
	 However, not all $\mathsf{QR}$- or SVD-based least-squares
	 fitting routines do this; for example, the widely-used
	 SVD-based routines given by
	 \cite[section~15.4]{Numerical-Recipes-2nd-edition}
	 do not perform such an scaling internally.
	 }
{}  I assign the individual data points the weights
$\bigl(\delta F_{\ell,\reg}^\internal\bigr)^{-2}$.


\subsection{Internal Error Estimates}
\label{sect-numerical-methods/internal-error-estimates}

I now consider the numerical computation of error estimates (bounds)
for the individual regularized self-force modes, the numerical force,
the tail force, and the overall self-force.  In this section I consider
only ``internal'' error estimates, those which can be computed from
(as part of) a single self-force calculation.  In
section~\ref{sect-numerical-methods/record-playback-error-estimates}
I consider ``record-playback'' error estimates derived from comparisons
between a low- and high-accuracy pair of self-force calculations,
and in section~\ref{sect-numerical-methods/actual-errors} I consider
``actual'' errors derived from comparisons between a self-force
calculation and a different and much more accurate calculation.


\subsubsection{Individual Regularized Self-Force Modes $F_{\ell,\reg}$}
\label{sect-numerical-methods/internal-error-estimates/individual-modes}

As noted in section~\ref{sect-Barack-Ori/self-force},
$F^{(+)}_{\ell,\reg} = F^{(-)}_{\ell,\reg}$.  However, due to the
finite-difference truncation errors in the numerical solution of
the wave equation~\eqref{eqn-wave}, the numerically-computed values
of~$F^{(+)}_{\ell,\reg}$ and~$F^{(-)}_{\ell,\reg}$ will differ
slightly.  I use this difference to derive an error estimate
(more accurately, an error \emph{bound}) for each regularized
self-force mode,
\begin{eqnarray}
F_{\ell,\reg} \pm \delta F_{\ell,\reg}^\internal
\hspace*{-5em}
	&        &						\nonumber\\
	& \eqdef &
		\thalf \left(
		       F^{(+)}_{\ell,\reg} + F^{(-)}_{\ell,\reg}
		       \right)
		\pm
		\thalf \left|
		       F^{(+)}_{\ell,\reg} - F^{(-)}_{\ell,\reg}
		       \right|
								      \,\text{.}
			     \label{eqn-individual-mode-internal-error-estimate}
\end{eqnarray}

[Notice that this internal error estimate does \emph{not} depend
in any way on the use of an AMR algorithm to solve the
wave equation~\eqref{eqn-wave}: $F^{(+)}_{\ell,\reg}$ and
$F^{(-)}_{\ell,\reg}$ would both still be well-defined even in a
unigrid simulation, and their difference would still be a measure
of the finite-differencing errors.

In section~\ref{sect-results/validation-of-error-estimates/individual-modes}
I present numerical evidence that these error estimates provide
reasonable (in fact, somewhat conservative) estimates of the actual
numerical errors in the individual regularized self-force modes.


\subsubsection{The Numerical Force $F_{\self,\num}$}
\label{sect-numerical-methods/internal-error-estimates/numerical-force}

The propagation of the individual regularized self-force modes' error
estimates~\eqref{eqn-individual-mode-internal-error-estimate} through
the numerical-force computation~\eqref{eqn-self-force/numerical-force}
is non-trivial, because we don't \textit{a priori} know whether or
to what extent the actual errors in the individual regularized
self-force modes~$F_{\ell,\reg}$ for different~$\ell$ are correlated.
(Since all the modes are calculated using the same basic numerical
scheme, some degree of correlation in their errors would not be
implausible.)

To investigate this question, I consider two extreme cases for how
the numerical force's error estimate $\delta F_\self^\internal$ might
be defined:
\begin{subequations}
			     \label{eqn-numerical-force-internal-error-estimate}
\begin{itemize}
\item	If the actual numerical errors in different modes are
	statistically independent, then the individual modes' error
	estimates~\eqref{eqn-individual-mode-internal-error-estimate}
	should be added in quadrature,
	\begin{equation}
	\left( \delta F_{\self,\num}^\internal \right)^2
		\eqdef	\sum_{\ell=0}^K
			\left( \delta F_{\ell,\reg}^\internal \right)^2
	      \label{eqn-numerical-force-internal-error-estimate/quadrature-sum}
	\end{equation}
\item	If the actual numerical errors in different modes are
	perfectly correlated, then the individual modes' error
	estimates~\eqref{eqn-individual-mode-internal-error-estimate}
	should be added arithmetically,
	\begin{equation}
	\delta F_{\self,\num}^\internal
		\eqdef \sum_{\ell=0}^K \delta F_{\ell,\reg}^\internal
	      \label{eqn-numerical-force-internal-error-estimate/arithmetic-sum}
	\end{equation}
	Barack and Sago~\cite{Barack-Sago-2010} use this formula
	to compute the numerical-force error given the (record-playback)
	error estimates of the individual modes.
\end{itemize}
\end{subequations}

In section~\ref{sect-results/validation-of-error-estimates/numerical-force}
I present numerical evidence that the actual numerical errors in
different modes are, if not completely independent, then at least
weakly enough correlated that the quadrature
sum~\eqref{eqn-numerical-force-internal-error-estimate/quadrature-sum}
provides a reliable error estimate for the numerical force, whereas
the arithmetic
sum~\eqref{eqn-numerical-force-internal-error-estimate/arithmetic-sum}
systematically overestimates the errors in the numerical force by a
factor of~$\sim\! 3$.


\subsubsection{The Tail Force $F_{\self,\tail}$}
\label{sect-numerical-methods/internal-error-estimates/tail-force-fitted}

For fitting the large-$\ell$ series expansion~\eqref{eqn-tail-series}
and computing the self-force tail force$F_{\self,\tail}$
via~\eqref{eqn-tail-force}, the same issue of statistical independence
versus correlation of the actual errors in $F_{\ell,\reg}$ for
different~$\ell$ arises again:
\begin{subequations}
				    \label{eqn-tail-force-internal-error-estimate}
\begin{itemize}
\item	If the actual numerical errors in different modes are
	statistically independent (and the individual modes' error
	estimates~\eqref{eqn-individual-mode-internal-error-estimate}
	are treated as the standard deviations of Gaussian distributions),
	then the standard theory of linear least-squares fitting
	can be applied (\cite{Lawson-Hanson-1974:least-squares,
Hammarling-1985:SVD-in-multivariate-statistics};
\cite[section~15.4]{Numerical-Recipes-2nd-edition}).
	The tail fit then provides the covariance matrix
	$\C$ for the fitted coefficients~$\{c_p\}$.
\footnote{
	 If the statistical assumptions don't actually hold,
	 $\C$ is more accurately termed the \emph{formal}
	 covariance matrix.
	 }
{}	Since the tail force can be written as the linear
	combination~\eqref{eqn-tail-force} of the fitted
	coefficients~$\{c_p\}$ with linear-combination
	coefficients~$\{\Gamma_p\}$, it's then easy to compute
	the estimated uncertainty in the tail force,
	\begin{equation}
	\left( \delta F_{\self,\tail}^\internal \right)^2
		\eqdef	\sum_{p,q} \C_{pq} \Gamma_p \Gamma_q
			\label{eqn-tail-force-internal-error-estimate/statistical}
	\end{equation}
	where the sum is over only the linear-combination
	coefficients $\{\Gamma_p\}$ corresponding to the fitted
	coefficients $\{c_p\}$.
\item	Alternatively, I can approximate the worst-case errors in the
	tail force~\eqref{eqn-tail-force} without assuming anything about
	the statistical independence of the actual numerical errors in
	different modes, as follows:  Suppose the set of $\ell$ for
	which~$F_{\ell,\reg}$ is used in the tail fit is
	$\{\ell_1, \ell_2, \ell_3, \dots, \ell_Q\}$.
	For each $k \in \{1, 2, 3, \dots, Q\}$, suppose
	$\eta_k \in \{-1,0,+1\}$, and define
	$F_{\ell_k,\reg}^\trial
		= F_{\ell_k,\reg} + \eta_k \, \delta F_{\ell_k,\reg}^\internal$.
	Then for each of the $3^Q$~possible combinations of
	$\eta_1$, $\eta_2$, $\eta_3$, \dots, $\eta_Q$, I perform
	a separate ``trail'' tail fit of the series
	expansion~\eqref{eqn-tail-series} to all the
	$F_{\ell_k,\reg}^\trial$, and compute the corresponding
	tail force$F_{\self,\tail}^\trial$ via~\eqref{eqn-tail-force}.
	Finally, I take the extreme range of these tail forces
	$F_{\self,\tail}^\trial$ among all $3^Q$~trial tail fits as a
	worst-case error estimate~$\delta F_{\self,\tail}^\internal$
	for the tail force $F_{\self,\tail}$,
	\begin{equation}
	\delta F_{\self,\tail}^\internal
		\eqdef \max_{\eta_1,\dots,\eta_Q}
		       \left| F_{\self,\tail}^\trial - F_{\self,\tail} \right|
								      \,\text{.}
			 \label{eqn-tail-force-internal-error-estimate/worst-case}
	\end{equation}
\end{itemize}
\end{subequations}

In section~\ref{sect-results/validation-of-error-estimates/tail-force}
I present numerical evidence that the statistical error
estimate~\eqref{eqn-tail-force-internal-error-estimate/statistical}
is moderately conservative, overestimating the actual numerical
errors in the tail force $F_{\self,\tail}$ by a factor of $\sim\! 2$,
while the worst-case
error estimate~\eqref{eqn-tail-force-internal-error-estimate/worst-case}
overestimates the actual errors by a factor of~$\sim\! 5$.


\subsubsection{Overall Self-Force}
\label{sect-numerical-methods/internal-error-estimates/self-force}

Given the internal error
estimates~\eqref{eqn-numerical-force-internal-error-estimate/quadrature-sum}
and~\eqref{eqn-tail-force-internal-error-estimate/statistical} for
the numerical force and tail force respectively, the question of their
statistical independence or lack thereof arises once again when computing
the overall self-force~$F_\self$ via~\eqref{eqn-self-force/numerical+tail}.
I again consider two possible choices for an internal error estimate
$\delta F_\self^\internal$ for $F_\self$:
\begin{subequations}
				  \label{eqn-self-force-internal-error-estimate}
\begin{itemize}
\item	As a best-case estimate (errors perfectly independent),
	I take the quadrature sum
	\begin{equation}
	\left( \delta F_\self^\internal \right)^2
		= \left( \delta F_{\self,\num}^\internal \right)^2
		  +
		  \left( \delta F_{\self,\tail}^\internal \right)^2
		   \label{eqn-self-force-internal-error-estimate/quadrature-sum}
	\end{equation}
\item	As a worst-case estimate (errors perfectly correlated),
	I take the arithmetic sum
	\begin{equation}
	\delta F_\self^\internal = \delta F_{\self,\num}^\internal
				   + \delta F_{\self,\tail}^\internal
		   \label{eqn-self-force-internal-error-estimate/arithmetic-sum}
	\end{equation}
\end{itemize}
\end{subequations}

In section~\ref{sect-results/validation-of-error-estimates/self-force}
I present numerical evidence that while both of these error estimates
are fairly reliable, the arithmetic-sum error
estimate~\eqref{eqn-self-force-internal-error-estimate/arithmetic-sum}
tends to give a slightly more accurate estimate of the actual errors
than the quadrature-sum error
estimate~\eqref{eqn-self-force-internal-error-estimate/quadrature-sum}.


\subsection{Record-Playback Error Estimates}
\label{sect-numerical-methods/record-playback-error-estimates}

To validate the internal error estimates, I pair each ``record'' AMR
solution of the wave equation~\eqref{eqn-wave} with a corresponding
``playback2'' numerical solution incorporating FMR of the recorded
grid structure by a factor of~$N_\fmr = 2$, in the manner discussed in
section~\ref{sect-numerical-methods/numerical-soln-of-wave-eqn/extended-fp-precision}.
My numerical code shows excellent 4th~order
convergence~\cite[figure~7]{Thornburg-2009:characteristic-AMR}, so
the finite-difference truncation errors in each playback2 evolution
are very close to a factor of $N_\fmr^4 = 16$~smaller than those
of the corresponding AMR record evolution.  I thus define the
``record-playback'' error estimate for each ``record'' regularized
self-force mode~$F_{\ell,\reg}$ by
\begin{equation}
\delta F_{\ell,\reg}^\rp
	\eqdef	\tfrac{16}{15}
		\left|
		F_{\ell,\reg}^{(\text{record})}
		- F_{\ell,\reg}^{(\text{playback2})}
		\right|
								      \,\text{.}
\end{equation}
[Here I'm implicitly assuming that finite-difference truncation errors
are the major contributor to the overall error in each $F_{\ell,\reg}$.
As discussed in
section~\ref{sect-numerical-methods/numerical-soln-of-wave-eqn/extended-fp-precision},
this is true in practice so long as the AMR error tolerance~$\AMRtolerance$
isn't too small, cf.~discussion in section~\ref{sect-results/tail-fits}.]

This same record-playback technique can also be used to estimate
the numerical errors in the ``record'' numerical force~$F_{\self,\num}$,
tail force~$F_{\self,\tail}$, and total self-force~$F_\self$,
\begin{subequations}
\begin{eqnarray}
\delta F_{\self,\num}^\rp
	& \eqdef &	\tfrac{16}{15}
			\left|
			F_{\self,\num}^{(\text{record})}
			- F_{\self,\num}^{(\text{playback2})}
			\right|
									\\
\delta F_{\self,\tail}^\rp
	& \eqdef &	\tfrac{16}{15}
			\left|
			F_{\self,\tail}^{(\text{record})}
			- F_{\self,\tail}^{(\text{playback2})}
			\right|
									\\
\delta F_\self^\rp
	& \eqdef &	\tfrac{16}{15}
			\left|
			F_\self^{(\text{record})}
			- F_\self^{(\text{playback2})}
			\right|
								      \,\text{.}
\end{eqnarray}
\end{subequations}

Notice that unlike the internal error estimates discussed in
section~\ref{sect-numerical-methods/internal-error-estimates},
which can be computed from a single numerical solution of the wave
equation~\eqref{eqn-wave}, the computation of any of these record-playback
error estimates requires a \emph{pair} of numerical solutions of
different accuracy (in this case, record and playback2); the
record-playback error estimate is only computed for the lower-accuracy
(in this case, record) member of the pair.


\subsection{Actual Errors}
\label{sect-numerical-methods/actual-errors}

Finally, for those particle orbits included in the highly accurate
frequency-domain self-force calculations of Detweiler, Messaritaki,
and Whiting~\cite{Detweiler-Messaritaki-Whiting-2003} and
Diaz-Rivera~\etal~\cite{Diaz-Rivera-etal-2004}, I can compute the
actual self-force errors as
\begin{equation}
\delta F_\self^\actual
	\eqdef	\left|
		F_\self - F_\self^{(\text{published})}
		\right|
					     \label{eqn-self-force-actual-error}
								      \,\text{.}
\end{equation}

Here I'm implicitly taking the published results as ``exact'',
i.e., I'm assuming that they're computed much more accurately than
my computations.  This is true for most, though not all, of the
numerical results presented in this paper.  In particular, for the
test case considered in section~\ref{sect-results}, Detweiler,
Messaritaki, and Whiting~\cite{Detweiler-Messaritaki-Whiting-2003}
give the self force as $F_\self = 1.378\,448\,28 \times 10^{-5} q^2/M$,
with an estimated uncertainty of $\Delta_\DMW = 2 \times 10^{-13} q^2/M$
(0.015~ppm).  I consider any ``actual errors''
defined by~\eqref{eqn-self-force-actual-error} which are less than
$3 \Delta_\DMW$ to be unreliable.


\subsection{Summary}
\label{sect-Barack-Ori/summary}

In summary, the numerical computation of the self force involves the
following steps:
\begin{enumerate}
\item	\label{step-solve-wave-eqn}
	Numerically solve the wave equation~\eqref{eqn-wave}
	for a suitable set of $(\ell,m)$, using either double or
	long-double floating-point arithmetic.  (All subsequent
	steps use double floating-point arithmetic.)
\item	\label{step-compute-F_ell,reg}
	Calculate the regularized self-force modes~$F_{\ell,\reg}$
	and their internal error estimates~$\delta F_{\ell,\reg}^\internal$
	for the corresponding set of~$\ell$,
	using~\eqref{eqn-F_ell-defn}, \eqref{eqn-F_self-pm},
	and~\eqref{eqn-individual-mode-internal-error-estimate}.
\item	\label{step-compute-F_self,num}
	Calculate the numerical force $F_{\self,\num}$
	and its internal error estimate $\delta F_{\self,\num}^\internal$
	using~\eqref{eqn-self-force/numerical-force} and one of the
	definitions~\eqref{eqn-numerical-force-internal-error-estimate}.
\item	\label{step-do-tail-fit-compute-F_self,tail}
	Perform the tail fit to determine the coefficients~$\{c_p\}$,
	then calculate the tail force~$F_{\self,\tail}$ and its
	internal error estimate~$\delta F_{\self,\tail}^\internal$
	using~\eqref{eqn-tail-force}, \eqref{eqn-tail-force-Gamma}, and
	one of the definitions~\eqref{eqn-tail-force-internal-error-estimate}.
\item	\label{step-compute-F_self}
	Compute the self force $F_\self$ and its internal error
	estimate~$\delta F_\self^\internal$
	using~\eqref{eqn-self-force/numerical+tail} and one of the
	definitions~\eqref{eqn-self-force-internal-error-estimate}.
\end{enumerate}


\section{Numerical Results}
\label{sect-results}


\subsection{Numerical Parameters}
\label{sect-results/numerical-params}

As a test case I take $r_p = 10M$; the particle's orbital period
is~$2\pi/\omega_p \approx 199\,M$.  In most cases I compute the
regularized self-force modes~$F_{\ell,\reg}$ and their internal error
estimates~$\delta F_{\ell,\reg}^\internal$
(step~\ref{step-compute-F_ell,reg} in the summary of
section~\ref{sect-Barack-Ori/summary}) for $0 \le \ell \le 30$,
$\ell = 35$, and $\ell = 40$.  Taking into account that
$a_{\ell m} = 0$ (and thus the wave equation~\eqref{eqn-wave} is
trivial) if $\ell - m$~is odd, this set of $\ell$ gives a total of
295 distinct~$(\ell,m)$ for which the wave equation~\eqref{eqn-wave}
must be solved numerically (step~\ref{step-solve-wave-eqn}).

To test the numerical computation over a wide range of cost/accuracy
tradeoffs, I perform steps~\ref{step-solve-wave-eqn}
and~\ref{step-compute-F_ell,reg} for each of the combinations of the
numerical-accuracy parameters (floating-point precision, AMR error
tolerance, and FMR refinement factor) shown with a ``\ok''~symbol
in table~\ref{tab-numerical-accuracy-params}.

For a given AMR error tolerance~$\AMRtolerance$, the
regularized self-force modes~$F_{\ell,\reg}$ have internal error
estimates~$\delta F_{\ell,\reg}^\internal$ which vary over almost
4~orders of magnitude over the range of $\ell$ I compute (this can
be seen in figure~\ref{fig-mode-internal-errors}).  This suggests that
a better ratio of accuracy to computational cost might be obtained
by applying a larger amount of FMR to those modes with the largest
internal error estimates, and a smaller amount of FMR (or none at all)
for those modes with relatively small internal error estimates.
For the long-double $\AMRtolerance = 10^{-19}$ calculation, I thus
also perform a further ``playback23'' calculation which uses FMR by
a factor of~$3$ for $12 \le \ell \le 15$, $22 \le \ell \le 30$,
$\ell = 35$, and $\ell = 40$ (these are shown with a ``(\ok)''~symbol
in table~\ref{tab-numerical-accuracy-params}), and FMR by a factor
of~$2$ for the other~$\ell$.

\begin{table}[bp]
\begin{center}
Numerical-Accuracy Parameters						\\[1ex]
\renewcommand{\arraystretch}{1.333}
\begin{ruledtabular}
\begin{tabular}{llcccccc}
Floating-Point	&
		& \multicolumn{6}{c}{FMR Refinement Factor}		\\
\cline{3-8}								\\[-2ex]
Precision	& $\AMRtolerance$
		& 1	& 2	& 3	& 4	& 6	& 8		\\
\hline									\\[-2ex]
double		& $10^{-14}$
		& \okrp	& \ok	&	&	&	&		\\
double		& $10^{-15}$
		& \okrp	& \okrp	& \okrp	& \shadecell
					  \ok	&	&		\\
double		& $10^{-16}$
		& \okrp	& \okrp	& \shadecell
				  \okrp	& \shadecell
					  \okrp	& \shadecell
						  \okrp	& \shadecell
							  \ok		\\[1ex]
\hline									\\[-3ex]
long-double	& $10^{-14}$
		& \okrp	& \ok	&	&	&	&		\\
long-double	& $10^{-15}$
		& \okrp	& \ok	&	&	&	&		\\
long-double	& $10^{-16}$
		& \okrp	& \ok	&	&	&	&		\\
long-double	& $10^{-17}$
		& \okrp	& \ok	&	&	&	&		\\
long-double	& $10^{-18}$
		& \okrp	& \ok	&	&	&	&		\\
long-double	& $10^{-19}$
		& \okrp	& \okrp	&(\ok)	&	&	&		\\
\end{tabular}
\end{ruledtabular}
\end{center}
\begin{flushleft}
\renewcommand{\arraystretch}{1.333}
\begin{tabular}{ccl}
\ok	& $=$	& $0 \le \ell \le 30$, $\ell = 35$, $\ell = 40$		\\
\okrp	& $=$	& $0 \le \ell \le 30$, $\ell = 35$, $\ell = 40$;	\\[-1ex]
	&	& record-playback error estimate can be computed	\\
(\ok)	& $=$	& $12 \le \ell \le 15$, $22 \le \ell \le 30$,
		  $\ell = 35$, $\ell = 40$				\\
\shadecell
	& $=$	& solution of the wave equation~\eqref{eqn-wave}
		  is seriously						\\[-1ex]
	&	& affected by floating-point roundoff errors
\end{tabular}
\end{flushleft}
\vspace*{-4mm}
\caption[Summary of Accuracy Parameters]
	{
	This table lists the numerical-accuracy parameters
	(the floating-point precision, AMR error tolerance~$\AMRtolerance$,
	and FMR refinement factor), and $\ell$ for which
	I have numerically solved the wave equation~\eqref{eqn-wave}
	(step~\ref{step-solve-wave-eqn} in the summary of
	section~\ref{sect-Barack-Ori/summary}) and computed
	the regularized self-force modes~$F_{\ell,\reg}$
	(step~\ref{step-compute-F_ell,reg}).  (This latter
	computation always uses double floating-point precision.)
	The shaded cells mark parameters where the solution
	of the wave equation~\eqref{eqn-wave} is seriously
	affected by floating-point roundoff errors.  (Results from
	these parameters are plotted as the ``double (bad)''
	points in figure~\ref{fig-mode-errors-scatterplot}.)
	}
\label{tab-numerical-accuracy-params}
\end{table}

\begin{figure}[bp]
\begin{center}
\includegraphics[scale=1.10]{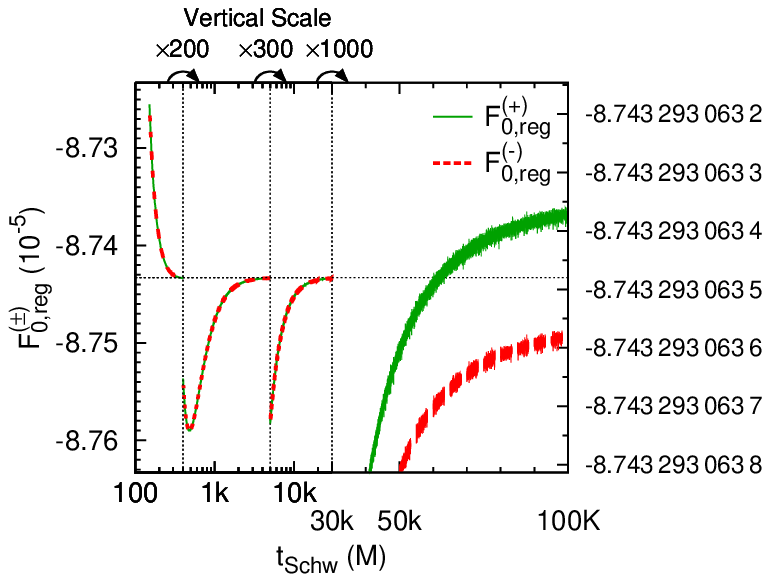}
\end{center}
\vspace{-10mm}
\begin{center}
\includegraphics[scale=1.10]{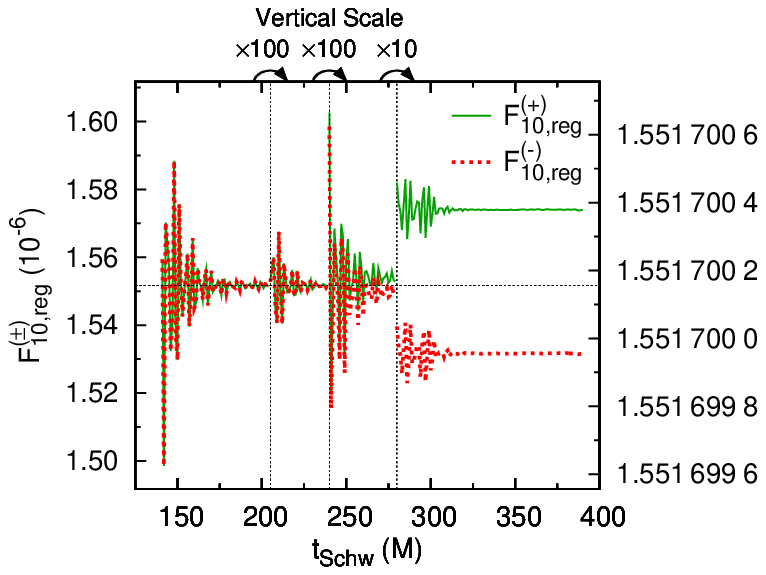}
\end{center}
\vspace{-5mm}
\caption[Examples of Individual Regularized Self-Force Modes' Time Decay]
	{
	[Color online]
	This figure shows the decay of the regularized self-force
	modes $F^{(\pm)}_{0,\reg}$~(top) and $F^{(\pm)}_{10,\reg}$~(bottom)
	towards their late-time values, for the $\AMRtolerance = 10^{-19}$
	``record'' evolution.  The vertical scale changes at each
	vertical dashed line, zooming in from left to right by the
	factors shown above each plot.  (The changing vertical scale
	also accounts for the apparent increase in the difference
	$F^{(+)}_{\ell,\reg} - F^{(-)}_{\ell,\reg}$ at later times;
	this difference is actually almost time-independent.)
	In the $F^{(\pm)}_{0,\reg}$ plot the horizontal scale
	changes from logarithmic for $t_\Schw \le 30\,000M$
	(shown as the upper of the two rows of time labels below the plot)
	to linear for $t_\Schw \ge 30\,000M$ (shown as the lower
	of the two rows of time labels below the plot).
	}
\label{fig-mode-time-decay}
\end{figure}

The required problem domain size for the numerical solution of the
wave equation~\eqref{eqn-wave} is set by how long it takes the
incorrect-initial-data perturbation to decay below the numerical error
level.  This size needs to be larger for smaller $\ell$ (where the
perturbation decays more slowly) and for greater accuracy (smaller
AMR error tolerance~$\AMRtolerance$ and/or larger FMR refinement factor).
For example, figure~\ref{fig-mode-time-decay} shows the time dependence
of $F^{(\pm)}_{0,\reg}$ (where the perturbation decays very slowly)
and of $F^{(\pm)}_{10,\reg}$ (where the perturbation decays fairly
rapidly); this latter case is is qualitatively similar to those of
the other $F^{(\pm)}_{\ell,\reg}$ with $\ell > 0$.  Based on trial
experiments with different problem-domain sizes, I have adopted the
problem-domain sizes given in table~\ref{tab-problem-domain-sizes}.

\begin{table}[bp]
\begin{center}
Problem-Domain Sizes							\\[1ex]
\begin{ruledtabular}
\begin{tabular}{lccrc}
$\ell$	& AMR Error Tolerance $\AMRtolerance$
			& \multicolumn{3}{c@{}}{Problem-Domain Size ($M$)}
									\\
\hline									\\[-2ex]
0	& $10^{-14}$, $10^{-15}$, $10^{-16}$
			& &  $30\,000$	&				\\
0 	& $10^{-17}$, $10^{-18}$, $10^{-19}$
			& \hspace{3.5em}
			  & $100\,000$	&				\\
1	&		& &   $5\,000$	&				\\
2	&		& & $1\,000$	&				\\
3--4	&		& &    $500$	&				\\
5--30	&		& &    $400$	&				\\
35	&		& &    $400$	&				\\
40	&		& &    $400$	&				
\end{tabular}
\end{ruledtabular}
\end{center}
\caption[Problem-Domain Sizes]
	{
	This table shows the problem-domain size used
	for each numerical evolution.
	}
\label{tab-problem-domain-sizes}
\end{table}

Because of the very slow decay of $\ell=0$ perturbations in
Schwarzschild spacetime (this is visible in figure~\ref{fig-mode-time-decay}),
I use very large problem-domain sizes for $\ell=0$ to ensure that
$F^{(\pm)}_{0,\reg}$ can be computed very accurately.  Since I use
an AMR numerical scheme~\cite{Thornburg-2009:characteristic-AMR}
where the numerical evolution's computational cost is strongly
concentrated near the particle worldline, even the very large
$\ell=0$ problem-domain sizes still only contribute a small fraction
(typically $\ltsim 15\%$) of the total computational cost of the
self-force calculation.
\footnote{
	 With an AMR scheme of this type, the total cost of a
	 numerical evolution at a given accuracy grows only
	 linearly with the problem-domain size, rather than
	 quadratically as would be the case for a unigrid scheme.
	 }
$^,$
\footnote{
	 For an AMR scheme such as mine there would be little benefit
	 in using a timelike inner boundary (and boundary condition)
	 of the type used by Haas~\cite{Haas-2007}: while this
	 could remove almost half of the problem domain, the
	 region removed would be distant from the particle worldline,
	 so its removal would only save a small part of the total
	 computational cost of the numerical evolution.
	 }

The problem-domain sizes shown in table~\ref{tab-problem-domain-sizes}
suffice to ensure that each $F_{\ell,\reg}$ is time-independent
to well within the numerical errors by the end of its numerical
evolution.  All the results reported here use $F^{(\pm)}_{\ell,\reg}$
values sampled $10M$ before the end of the evolution.

To further explore cost/accuracy tradeoffs in self-force calculations,
for each combination of numerical-accuracy parameters given in
table~\ref{tab-numerical-accuracy-params} I have computed the
numerical force $F_{\self,\num}$ (step~\ref{step-compute-F_self,num}
in the summary of section~\ref{sect-Barack-Ori/summary}),
performed the tail fit and computed the tail force $F_{\self,\tail}$
(step~\ref{step-do-tail-fit-compute-F_self,tail}),
and computed the self-force (step~\ref{step-compute-F_self})
for each for the numerical-force and tail-fit parameters shown in
table~\ref{tab-numerical-force-and-tail-fit-params}.

\begin{table}[bp]
\begin{center}
Numerical-Force and Tail-Fit Parameters					\\[1ex]
\begin{ruledtabular}
\begin{tabular}{llcccc}
	&		& \multicolumn{2}{c}{Analytical $c_2$}
					& \multicolumn{2}{c}{Tail-Fitted $c_2$}
									\\
\cline{3-4} \cline{5-6}							\\[-2ex]
$K$	& $\{\ell\}$ in tail fit
			& $\{4\}$
				& $\{4,6\}$
					& $\{2,4\}$
						& $\{2,4,6\}$		\\[1ex]
\hline									\\[-2ex]
15	& 10--15	& \ok	& \ok	& \ok	& \ok			\\
20	& 15--20	&	& \ok	&	& \ok			\\
25	& 20--25	&	& \ok	&	& \ok			\\
30	& 20--30	&	& \ok	&	& \ok			\\
30	& 20--30, 35, 40
			&	& \ok	&	& \ok			\\
\end{tabular}
\end{ruledtabular}
\end{center}
\caption[Numerical-Force and Tail-Fit Parameters]
	{
	This table shows the numerical-force and tail-fit parameters.
	$K$ is the maximum~$\ell$ included in the numerical force.
	}
\label{tab-numerical-force-and-tail-fit-params}
\end{table}


\subsection{Overview of the Numerical Results}
\label{sect-results/overview}

\begin{figure}[bp]
\begin{center}
\hspace*{-1mm}
\includegraphics[scale=0.90]{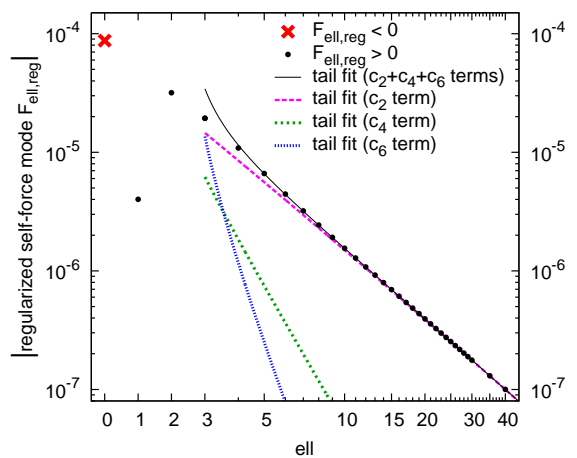}
\end{center}
\vspace{-7.5mm}
\caption[Overview of $F_{\ell,\reg}$]
	{
	[Color online]
	This figure shows the regularized self-force modes
	$F_{\ell,\reg}^\internal$.
	The $\ell$~scale is linear from $\ell = 0$ to~$3$,
	then logarithmic from $\ell = 3$ to~$40$.
	}
\label{fig-modes-overview}
\end{figure}

Figure~\ref{fig-modes-overview} shows the regularized self-force
modes~$F_{\ell,\reg}$ for the most accurate ``record'' evolution
($\AMRtolerance = 10^{-19}$).  Notice that the large-$\ell$ modes
are very closely approximated by the tail fit~\eqref{eqn-tail-series};
I discuss this further in section~\ref{sect-results/tail-fits}.

\begin{figure}[bp]
\begin{center}
\hspace*{-1mm}
\includegraphics[scale=0.90]{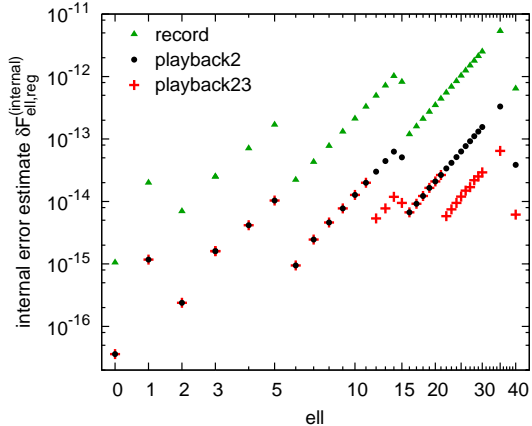}
\end{center}
\vspace{-7.5mm}
\caption[Overview of $\delta F_{\ell,\reg}^\internal$]
	{
	[Color online]
	This figure shows the regularized self-force modes'
	internal error estimates $\delta F_{\ell,\reg}^\internal$
	for the $\AMRtolerance = 10^{-19}$ evolutions.
	The $\ell$~scale is linear from $\ell = 0$ to~$3$,
	then logarithmic from $\ell = 3$ to~$40$.
	}
\label{fig-mode-internal-errors}
\end{figure}

Figure~\ref{fig-mode-internal-errors} shows the regularized self-force
modes' internal error estimates~$\delta F_{\ell,\reg}^\internal$ for
the record, playback2, and playback23 $\AMRtolerance = 10^{-19}$~evolutions.
As $\ell$ increases, the solutions of the wave equation~\eqref{eqn-wave}
oscillate more rapidly in space and time, so the finite-difference
truncation errors for any fixed numerical resolution increase rapidly.
The AMR algorithm responds to this by decreasing $\Delta vu_{\min}$
in discrete factor-of-$2$ steps, each of which decreases the global
finite-difference truncation error of the solution $\phi_{\ell m}$
by very close to a factor of $2^4 = 16$.  This accounts for the
``stepped'' appearance of the error estimates visible in
figure~\ref{fig-mode-internal-errors}.  (Notice also that as
intended, the playback23 error estimates show much less dynamic
range than the record or playback2 estimates.)

Depending on~$\ell$, the AMR algorithm uses between 5 and 8~levels of
2:1~mesh refinement for these evolutions.  For the record evolution,
the grid resolution of the finest refinement level varies from~$M/128$
to~$M/2048$ depending on $\ell$; the resolutions for the playback2
and playback23 evolutions are correspondingly finer.

The speedup factor of the AMR algorithm over an equivalent-resolution
(and thus roughly equivalent-accuracy) FMR evolution is typically
$30$ to $40$ for evolutions using an $\ell=0$ problem-domain size
of $30\,000 M$, and $200$ to $400$ for evolutions (such as the
$\AMRtolerance = 10^{-19}$ ones) using an $\ell=0$ problem-domain
size of $100\,000 M$.

\subsection{Validation of the Error Estimates}
\label{sect-results/validation-of-error-estimates}

In this section I present numerical tests to validate the internal
error estimates described in
section~\ref{sect-numerical-methods/internal-error-estimates}
against the record-playback error estimates described in
section~\ref{sect-numerical-methods/record-playback-error-estimates},
and (in those cases where the actual errors are known) to validate
the record-playback error estimates against the actual errors.  For
the comparisons of internal with record-playback error estimates I
use the results from all of the numerical-accuracy parameters for
which a record-playback error estimate can be computed (these
parameters are shown in table~\ref{tab-numerical-accuracy-params}).
To prevent inaccuracies in the tail fit from contaminating the error
estimates, in this section I consider only the highest-accuracy set
of numerical-force and tail-fit parameters shown in
table~\ref{tab-numerical-force-and-tail-fit-params} (i.e., those in
in the last row of the table): the numerical force sums modes up to
$K=30$, the tail fit includes the modes $20 \le \ell \le 30$, $\ell=35$,
and $\ell=40$, and the tail fit fits either $\{c_4,c_6\}$ (with $c_2$
given analytically by~\eqref{eqn-c_2-tail-fit-coeff}) or $\{c_2,c_4,c_6\}$.


\subsubsection{Individual Regularized Self-Force Modes $F_{\ell,\reg}$}
\label{sect-results/validation-of-error-estimates/individual-modes}

\begin{figure}[bp]
\begin{center}
\includegraphics[scale=1.10]{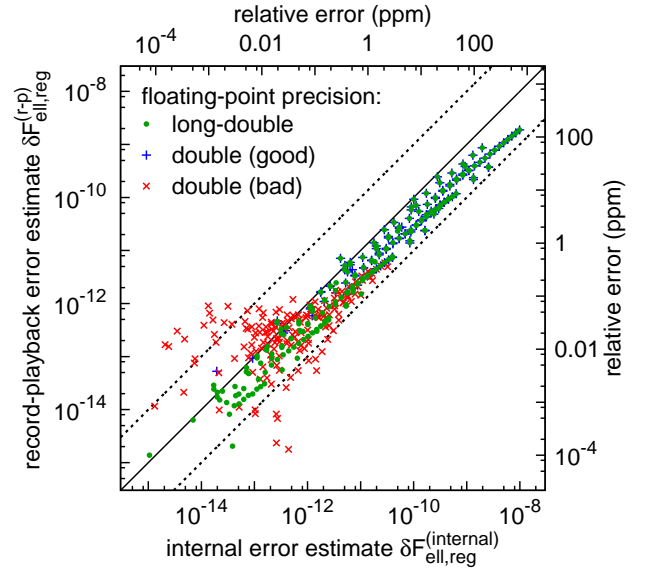}
\end{center}
\vspace{-7.5mm}
\caption[Scatterplot of Record-Playback versus Internal Error Estimates
	 for Individual Modes $F_{\ell,\reg}$]
	{
	[Color online]
	This figure shows a scatterplot of the record-playback
	error estimate~$\delta F_{\ell,\reg}^\rp$ versus the
	internal error estimate~$\delta F_{\ell,\reg}^\internal$.
	The solid and dashed lines show the cases where the
	record-playback error estimate is identical to or an order
	of magnitude larger/smaller than the internal error estimate,
	respectively.  The relative-error scales are relative to
	the overall self force $F_\self$.
	}
\label{fig-mode-errors-scatterplot}
\end{figure}

Figure~\ref{fig-mode-errors-scatterplot} shows a scatterplot of the
the record-playback error estimates $\delta F_{\ell,\reg}^\rp$ versus
the internal error
estimates~\eqref{eqn-individual-mode-internal-error-estimate}.
The internal error estimates~$\delta F_{\ell,\reg}^\internal$
for evolutions done in long-double floating-point precision are
generally generally consistent with the record-playback
error estimates~$\delta F_{\ell,\reg}^\rp$, and are somewhat
conservative (the internal error estimates tend to overestimate
the record-playback error estimates).

For evolutions done in double floating-point precision the internal
error estimates are similarly consistent and conservative for
the ``good'' parameters which are \emph{not} shown as shaded
in table~\ref{tab-numerical-accuracy-params}
However, for the ``bad'' parameters which are shown as shaded in
table~\ref{tab-numerical-accuracy-params} the internal error estimates
scatter widely about the record-playback error estimates, often
deviating by up to two orders of magnitude.
This is due to floating-point rounding errors contaminating the
numerical solution of the wave equation~\eqref{eqn-wave}
(step~\ref{step-solve-wave-eqn} in the summary of
section~\ref{sect-Barack-Ori/summary}).

\subsubsection{Self-Force Numerical Sum $F_{\self,\num}$}
\label{sect-results/validation-of-error-estimates/numerical-force}

\begin{figure}[bp]
\begin{center}
\includegraphics[scale=1.10]{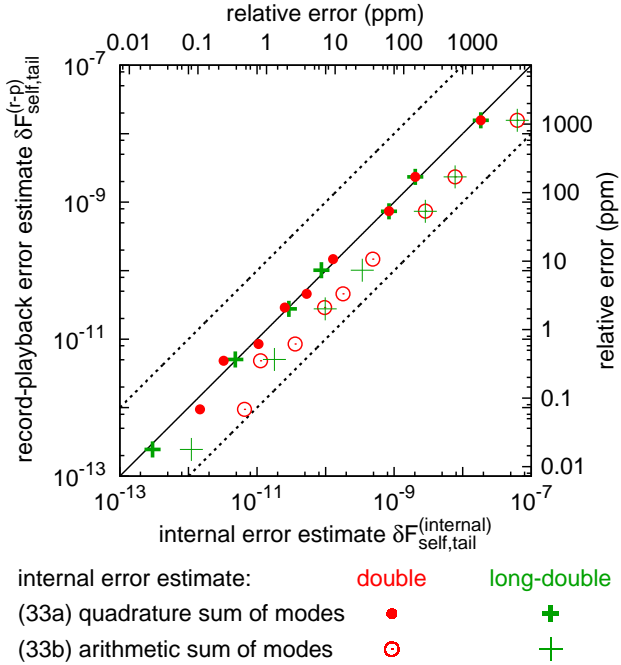}
\end{center}
\vspace{-7.5mm}
\caption[Scatterplot of Record-Playback versus Internal Error Estimates
	 for the Numerical Force]
	{
	[Color online]
	This figure shows a scatterplot of the numerical-force
	record-playback estimate~$\delta F_{\self,\num}^\rp$ versus
	the internal error estimate~$\delta F_{\self,\num}^\internal$
	(the latter computed using each of the
	definitions~\eqref{eqn-numerical-force-internal-error-estimate}).
	The solid and dashed lines show the cases where the
	record-playback error estimate is identical to or an order
	of magnitude larger/smaller than the internal error estimate,
	respectively.  The relative-error scales are relative to
	the overall self force $F_\self$.
	}
\label{fig-numerical-force-errors-scatterplot}
\end{figure}

Figure~\ref{fig-numerical-force-errors-scatterplot}
shows a scatterplot of the numerical-force record-playback
error estimate~$\delta F_{\self,\num}^\rp$ versus the internal error
estimate~$\delta F_{\self,\num}^\internal$, the latter computed using each
of the definitions~\eqref{eqn-numerical-force-internal-error-estimate}.
The arithmetic-sum internal error
estimate~\eqref{eqn-numerical-force-internal-error-estimate/arithmetic-sum}
systematically overestimates the record-playback error estimate by
a factor of~$\sim\! 3.5$.  In contrast, the quadrature-sum internal error
estimate~\eqref{eqn-numerical-force-internal-error-estimate/quadrature-sum}
is quite accurate.
\footnote{
	 Notice that the set of modes considered here \emph{includes}
	 the ``double (bad)'' regularized self-force modes plotted
	 in figure~\ref{fig-mode-errors-scatterplot} and discussed in
	 section~\ref{sect-results/validation-of-error-estimates/individual-modes}.
	 Evidently the averaging inherent in summing 33~of these
	 modes greatly reduces the effects of the floating-point
	 roundoff error contamination of the individual modes.
	 }
{}  Based on this, I adopt the quadrature-sum internal error
estimate~\eqref{eqn-numerical-force-internal-error-estimate/quadrature-sum}
hereinafter.

\subsubsection{Self-Force Tail Sum $F_{\self,\tail}$}
\label{sect-results/validation-of-error-estimates/tail-force}

\begin{figure}[bp]
\begin{center}
\includegraphics[scale=1.10]{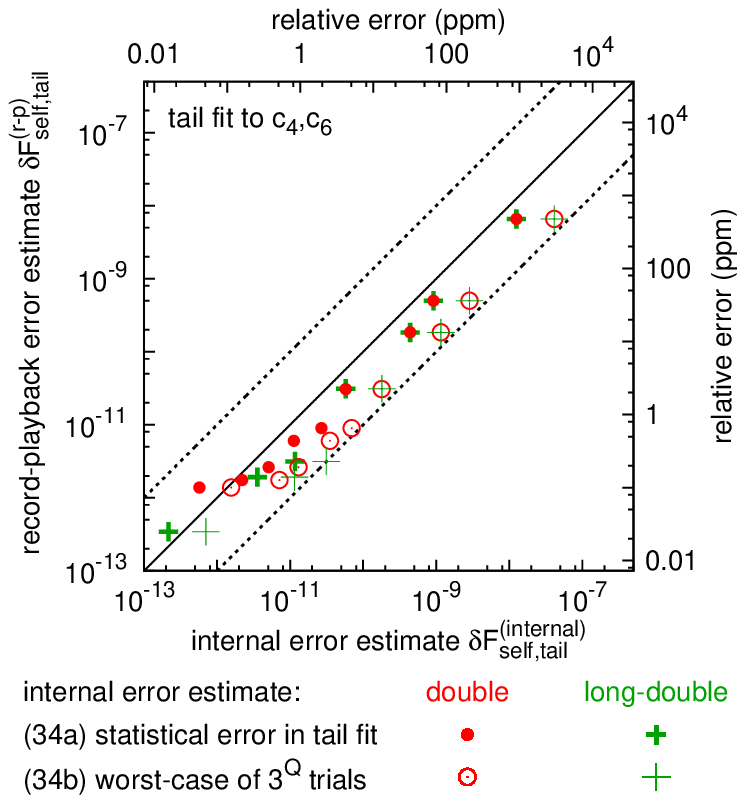}
\includegraphics[scale=1.10]{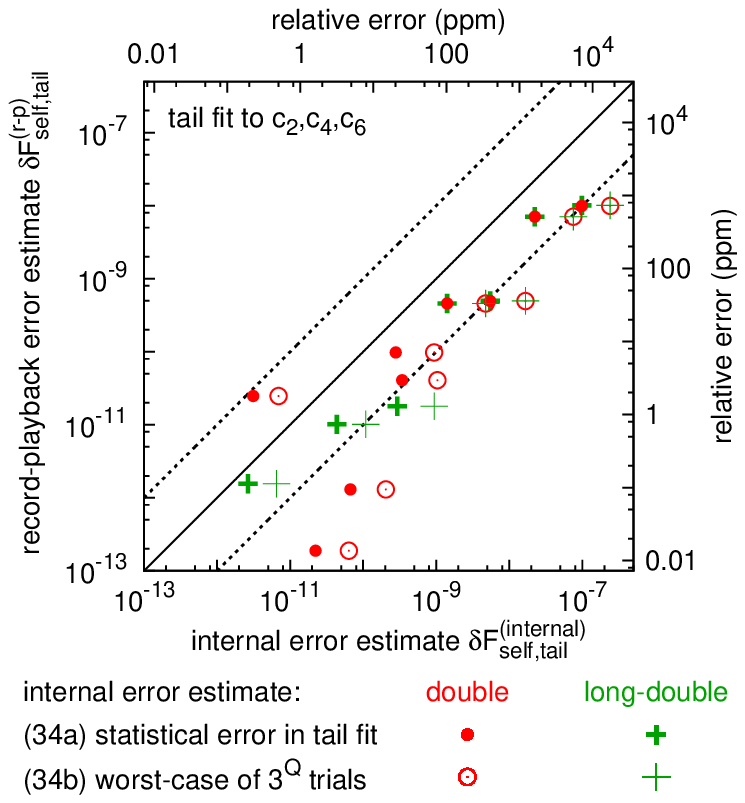}
\end{center}
\vspace{-7.5mm}
\caption[Scatterplot of Record-Playback versus Internal Error Estimates
	 for the Tail Sum]
	{
	[Color online]
	This figure shows a scatterplot of the tail-force
	record-playback estimate~$\delta F_{\self,\tail}^\rp$ versus
	the internal error estimate~$\delta F_{\self,\tail}^\internal$
	(the latter computed using each of the
	definitions~\eqref{eqn-tail-force-internal-error-estimate}),
	for tail fits to $\{c_4,c_6\}$~(top) and $\{c_2,c_4,c_6\}$~(bottom).
	The solid and dashed lines show the cases where the
	record-playback error estimate is identical to or an order
	of magnitude larger/smaller than the internal error estimate,
	respectively.  The relative-error scales are relative to
	the overall self force $F_\self$.
	}
\label{fig-tail-fit-errors-scatterplot}
\end{figure}

Figure~\ref{fig-tail-fit-errors-scatterplot} shows a scatterplot of
the tail-force record-playback error estimate~$\delta F_{\self,\tail}^\rp$
versus the internal error estimate~$\delta F_{\self,\tail}^\internal$
(the latter computed using each combination of the
definitions~\eqref{eqn-tail-force-internal-error-estimate}
and fitting either $\{c_4,c_6\}$ or $\{c_2,c_4,c_6\}$).
When fitting $\{c_4,c_6\}$ (with $c_2$ given analytically
by~\eqref{eqn-c_2-tail-fit-coeff}), for both double and long-double
floating-point precision the worst-case-of-$3^Q$-trials internal error
estimate~\eqref{eqn-tail-force-internal-error-estimate/worst-case}
tends to systematically overestimate the record-playback error estimate,
while the statistical internal error
estimate~\eqref{eqn-tail-force-internal-error-estimate/statistical}
is much more accurate.  Based on this, I adopt the statistical internal
error estimate~\eqref{eqn-tail-force-internal-error-estimate/statistical}
hereinafter.  For both precisions,
the internal error estimates change from
being overestimates to underestimates of the record-playback error
estimates for the two smallest-error points.

When fitting $\{c_2,c_4,c_6\}$, the long-double internal error
estimates are still consistent and somewhat conservative, but the
three smallest-error internal error estimates scatter widely about
the corresponding record-playback error estimates.  This appears to
be due to the ill-conditioning of the $\{c_2,c_4,c_6\}$ tail fit
(cf.~discussion in section~\ref{sect-numerical-methods/tail-fit},
particularly table~\ref{tab-tail-fit-basis-degeneracy}) amplifying
the floating-point rounding errors in the individual regularized
self-force modes $F_{\ell,\reg}$.

I discuss further ``quality control'' checks based on the tail fits'
$\chi^2$ and residuals in section~\ref{sect-results/tail-fits}.

\subsubsection{Overall Self-Force}
\label{sect-results/validation-of-error-estimates/self-force}

\begin{figure}[bp]
\begin{center}
\includegraphics[scale=1.10]{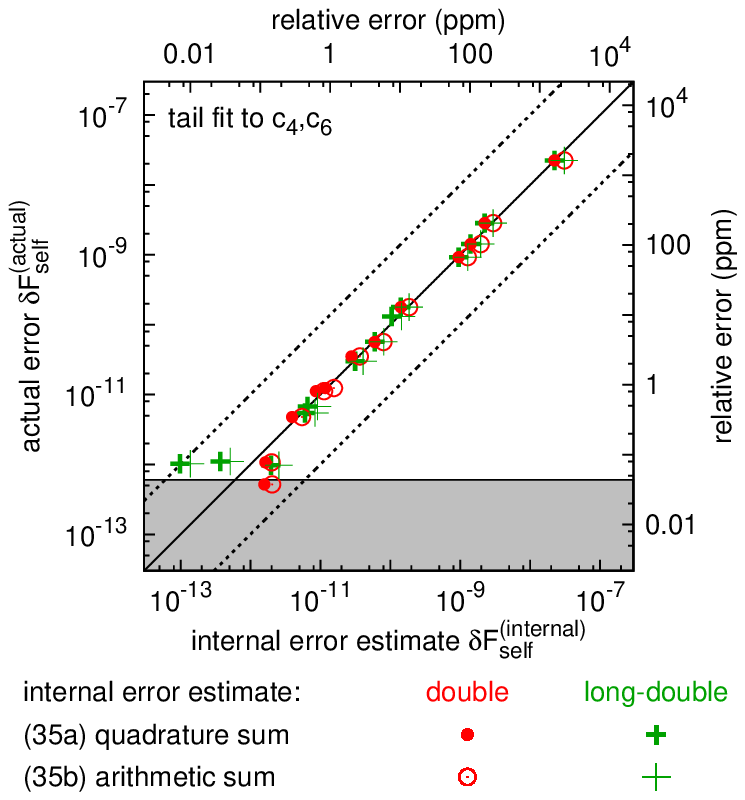}
\includegraphics[scale=1.10]{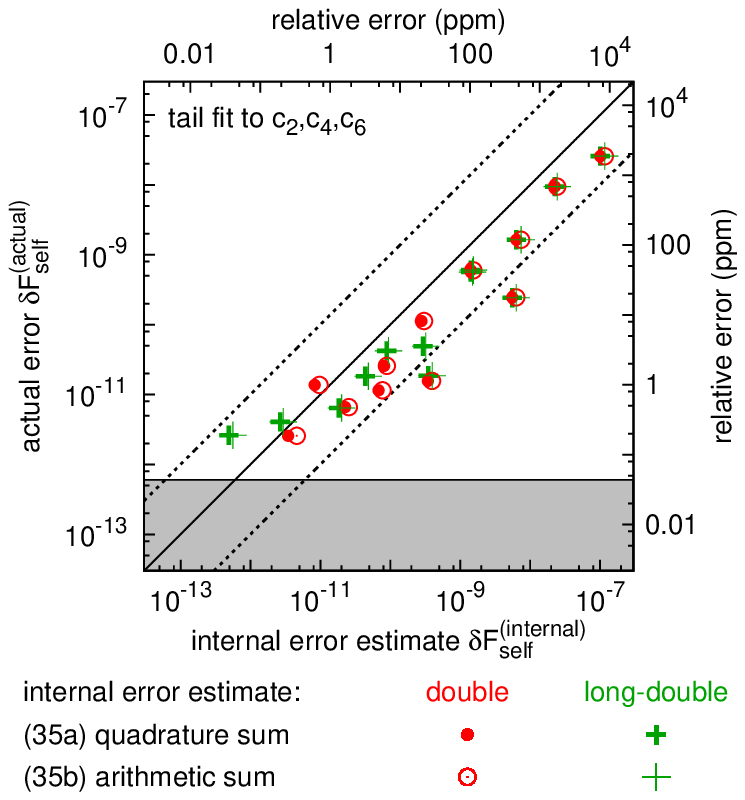}
\end{center}
\vspace{-5mm}
\caption[Scatterplot of Actual versus Internal Error Estimates
	 for the Self Force]
	{
	[Color online]
	This figure shows a scatterplot of the self-force
	actual errors~$\delta F_\self^\actual$ versus the
	internal error estimates~$\delta F_\self^\internal$
	(the latter computed using each of the
	definitions~\eqref{eqn-self-force-internal-error-estimate}),
	for tail fits to $\{c_4,c_6\}$~(top) and $\{c_2,c_4,c_6\}$~(bottom).
	The solid and dashed lines show the cases where the
	actual errors are identical to or an order of magnitude
	larger/smaller than the internal error estimates,
	respectively.  The actual-error values are unreliable
	in the shaded region of each plot.
	The relative-error scales are relative to
	the overall self force $F_\self$.
	}
\label{fig-self-force-errors-scatterplot}
\end{figure}

As discussed in section~\ref{sect-numerical-methods/actual-errors},
I can compute the actual error of the overall self-force by comparing
my calculations against previously published highly-accurate frequency-domain
calculations.  Figure~\ref{fig-self-force-errors-scatterplot} shows
a scatterplot of the actual self-force errors~$\delta F_\self^\actual$
versus the self-force internal error estimates~$\delta F_\self^\internal$
(the latter computed using each combination of the
definitions~\eqref{eqn-self-force-internal-error-estimate})
and for tail fits to $\{c_4,c_6\}$ or $\{c_2,c_4,c_6\}$).
All of the error estimates are fairly accurate.

\subsubsection{Record-Playback Error Estimates}
\label{sect-results/validation-of-error-estimates/record-playback}

\begin{figure}[bp]
\begin{center}
\includegraphics[scale=1.10]{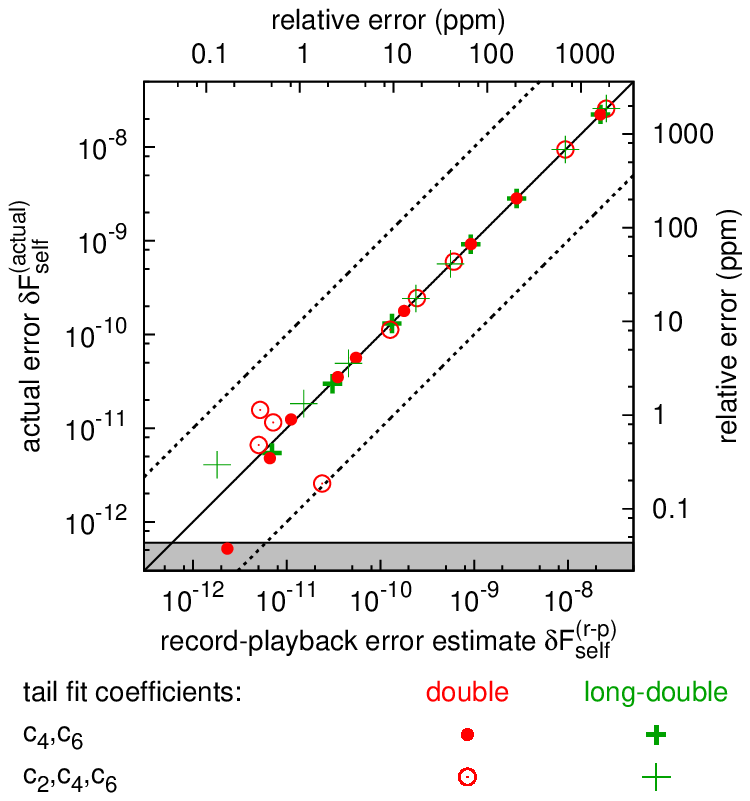}
\end{center}
\vspace{-5mm}
\caption[Scatterplot of Actual versus Record-Playback Error Estimates
	 for the Self Force]
	{
	[Color online]
	This figure shows a scatterplot of the self-force
	actual errors~$\delta F_\self^\actual$ versus the
	record-playback error estimates~$\delta F_\self^\rp$.
	The solid and dashed lines show the cases where the
	actual errors are identical to or an order of magnitude
	larger/smaller than the record-playback error estimates,
	respectively.   The actual-error values are unreliable
	in the shaded region of each plot.
	The relative-error scales are relative to
	the overall self force $F_\self$.
	}
\label{fig-rp-errors-scatterplot}
\end{figure}

The self-force actual errors can also be used to validate the
record-playback error estimates.  Figure~\ref{fig-rp-errors-scatterplot}
shows a scatterplot of the actual self-force errors~$\delta F_\self^\actual$
versus the record-playback error estimates~$\delta F_\self^\rp$.
The actual errors are very similar to the record-playback error
estimates.

\subsection{The Tail Fits}
\label{sect-results/tail-fits}

The quality of a tail fit can be assessed via the fit's $\chi^2$:
if $\chi^2$ lies outside the ``plausible'' range
$[\chi^2_{2.5\%}, \chi^2_{97.5\%}]$, where $\chi^2_{\beta\%}$ is the
$\beta\%$~percentile of the $\chi^2$~distribution for the appropriate
number of degrees of freedom (here $11$~for fitting~$\{c_4,c_6\}$,
or $10$~for fitting~$\{c_2,c_4,c_6\}$), we can reject the null hypothesis
that the the nonzero fit residuals are solely due to (independent
Gaussian-distributed) random errors of magnitudes given by the
individual modes' internal error estimates~$\delta F_{\ell,\reg}^\internal$.
A corollary of this is that the statistical tail-fit error
estimate~\eqref{eqn-tail-force-internal-error-estimate/statistical}
becomes unreliable.

\begin{figure}[bp]
\begin{center}
\includegraphics[scale=1.10]{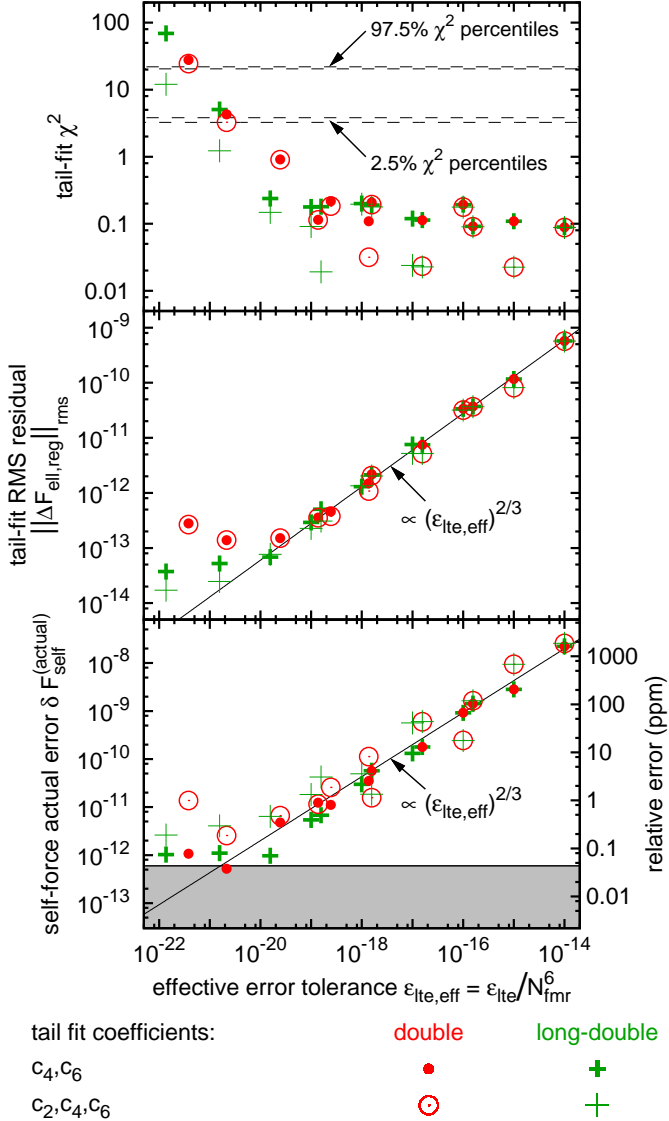}
\end{center}
\vspace{-7.5mm}
\caption[Tail-Fit $\chi^2$ and Residuals]
	{
	[Color online]
	This figure shows the $\chi^2$ (upper plot),
	RMS residuals~$\big\| \Delta F_{\ell,\reg} \big\|_\rms$ (middle plot),
	and self-force actual error~$\delta F_\self^\actual$ (lower plot)
	for each set of numerical-accuracy parameters listed in
	table~\ref{tab-numerical-accuracy-params}.
	In the $\chi^2$~plot, the two pairs of dashed lines show the
	$2.5\%$ and $97.5\%$~percentiles of the $\chi^2$~distribution
	for $11$~degrees of freedom (appropriate for fitting $\{c_4,c_6\}$)
	and for $10$~degrees of freedom (appropriate for fitting
	$\{c_2,c_4,c_6\}$).  In the RMS-residual plot the diagonal
	line shows the $\ErrorToleranceEffective^{2/3}$ scalings
	expected for my characteristic AMR
	algorithm~\cite{Thornburg-2009:characteristic-AMR}.
	In the actual-error plot, the actual-error values
	are unreliable in the shaded region.
	}
\label{fig-tail-fit-Chi2-residuals-errors}
\end{figure}

Figure~\ref{fig-tail-fit-Chi2-residuals-errors} shows the tail-fit $\chi^2$
for fitting both $\{c_4,c_6\}$ and $\{c_2,c_4,c_6\}$ for each set of
numerical-accuracy parameters listed in
table~\ref{tab-numerical-accuracy-params}.  For effective error tolerances
$\ErrorToleranceEffective \gtsim 10^{-20}$ the fits are very well-behaved:
$\chi^2$ is small,
\footnote{
	 In fact, $\chi^2 \ll \chi^2_{2.5\%}$.  This is due to
	 the individual regularized self-force modes'
	 internal error estimates~$\delta F_{\ell,\reg}^\internal$
	 systematically overrestimating the actual numerical errors
	 (cf.~figure~\ref{fig-mode-errors-scatterplot} and discussion in
	 section~\ref{sect-results/validation-of-error-estimates/individual-modes}).
	 }
{} and both the RMS residuals~$\big\| \Delta F_{\ell,\reg} \big\|_\rms$
and the self-force actual errors $\delta F_\self^\actual$ decrease
$\propto \ErrorToleranceEffective^{2/3}$ as $\ErrorToleranceEffective$
decreases, as expected for a characteristic evolution scheme with
4th/6th~order global/local finite differencing accuracy.

However, for $\ErrorToleranceEffective \ltsim 10^{-20}$ the fits show
several undesirable characteristics: $\chi^2$ increases, the RMS residuals
either begin to increase with decreasing $\ErrorToleranceEffective$
(double floating-point precision) or decrease at a slower rate than
$\propto \ErrorToleranceEffective^{2/3}$ (long-double floating-point
precision), and the actual errors either increase for the very smallest
$\ErrorToleranceEffective$ (double floating-point precision), level
off as $\ErrorToleranceEffective$ decreases (long-double floating-point
precision, tail fit to $\{c_4,c_6\}$ only), or decrease at a slower rate
than $\propto \ErrorToleranceEffective^{2/3}$ (long-double floating-point
precision, tail fit to $\{c_2,c_4,c_6\}$.  These effects are due to
floating-point roundoff errors contaminating the various steps of the
calculation.

\subsection{Cost/Accuracy Tradeoffs}
\label{sect-results/cost-accuracy-tradeoffs}

There are a number of cost/accuracy tradeoffs inherent in the choice
of the various numerical parameters in the self-force computation.

The computational cost is overwhelmingly dominated by the
numerical solution of the wave equation~\eqref{eqn-wave}
(step~\ref{step-solve-wave-eqn} in the summary of
section~\ref{sect-Barack-Ori/summary}), and is determined by
the combination of the set of~$\ell$ for which the wave
equation~\eqref{eqn-wave} is solved, and the problem-domain sizes,
floating-point precision, AMR error tolerances~$\AMRtolerance$,
and FMR refinement factors (if any) used in that solution.

In general, the problem-domain size should be chosen just large
enough to render the errors induced by the remaining time dependence
of $F_{\ell,\reg}$ small in comparison to other numerical errors.
The problem-domain size required to ensure this varies with the
magnitude of the other numerical errors, being larger for higher
accuracies (smaller errors).  The required problem-domain size
also varies strongly with $\ell$, being much larger for small~$\ell$
(cf.~discussion in section~\ref{sect-results/numerical-params}).
For simplicity, in this work I have only adjusted the problem-domain
sizes at the very coarse level shown in table~\ref{tab-problem-domain-sizes}.
A more careful adjustment would substantially improve the efficiency
of the computation.

For my computational scheme, AMR and FMR are of almost equal
efficiencies.  That is, the grid structure, computational cost,
and accuracy attained from an evolution using AMR with error
tolerance~$\AMRtolerance$ and FMR by a refinement factor of~$N_\fmr$
are all very similar to those obtained from a purely-AMR evolution
using the corresponding effective error
tolerance~$\ErrorToleranceEffective \eqdef \AMRtolerance / N_\fmr^6$.
Assuming that the problem-domain sizes are large enough so that
the remaining time dependence of $F_{\ell,\reg}$ isn't a significant
contributor to the overall error budget, the parameter space
for cost-performance tradeoffs in numerically solving the
wave equation~\eqref{eqn-wave} can thus be simplified to just
the effective error tolerance~$\ErrorToleranceEffective$.

The accuracy of a self-force computation is then determined by the
combination of the set of~$\ell$ for which the wave equation~\eqref{eqn-wave}
is solved, the effective error tolerance~$\ErrorToleranceEffective$
of this solution, $K$~(the maximum~$\ell$ included in the numerical force),
the set of $\ell$ used in fitting the tail series~\eqref{eqn-tail-series},
and the set of orders~$p$ and corresponding coefficients $\{c_p\}$
included in this series.  This is a large parameter space; for
present purposes I restrict consideration to those parameter
combinations listed in tables~\ref{tab-numerical-accuracy-params}
and~\ref{tab-numerical-force-and-tail-fit-params}.
 
\begin{figure}[bp]
\begin{center}
\includegraphics[scale=1.10]{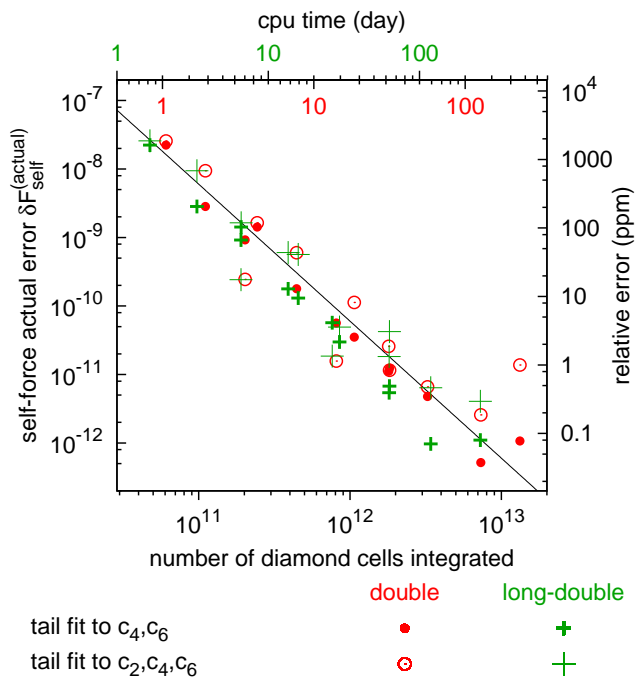}
\end{center}
\vspace{-7.5mm}
\caption[Cost/Accuracy Tradeoffs]
	{
	[Color online]
	This figure shows an overview of the cost-accuracy
	tradeoffs for the highest-accuracy set of numerical-force
	and tail-fit parameters shown in
	table~\ref{tab-numerical-force-and-tail-fit-params}.
	Notice that there are two different CPU-time scales:
	the upper (outer) scale is for evolutions in long-double
	floating-point precision, while the lower (inner) scale
	is for evolutions in double floating-point precision.
	The solid line shows the expected scaling
	$\delta F_\self^\actual \propto N^{-2}$.
	}
\label{fig-cost-accuracy}
\end{figure}

For present purposes, it's useful to quantify the computational cost
of an evolution by the total number of diamond cells integrated by the
AMR algorithm.  This is closely proportional to the overall CPU time
used, with the constant of proportionality (the CPU time per diamond
cell) being about $1.5$~($3.0$)~microseconds per diamond cell for
double (long-double) floating-point precision on the processors used
here.  Figure~\ref{fig-cost-accuracy} gives an overview of the
cost-accuracy tradeoffs for the highest-accuracy set of
numerical-force and tail-fit parameters shown in
table~\ref{tab-numerical-force-and-tail-fit-params} (i.e., those in
in the last row of the table): the numerical force sums modes up to
$K=30$, the tail fit includes the modes $20 \le \ell \le 30$, $\ell=35$,
and $\ell=40$, and the tail fit fits either $\{c_4,c_6\}$ (with $c_2$
given analytically by~\eqref{eqn-c_2-tail-fit-coeff}) or $\{c_2,c_4,c_6\}$.
It should be noted that the costs shown in this figure are for
computations with very conservative (large) problem-domain sizes,
and the wave equation~\eqref{eqn-wave} solve for a large set of
$\ell$.  The costs could be greatly reduced with only a minor impact
on the accuracy by using smaller problem-domain sizes and a smaller
set of $\ell$.

\subsection{Results for the Self-Force}
\label{sect-results/self-force}

Tables~\ref{tab-self-force-and-error-results/fit-order46}
and~\ref{tab-self-force-and-error-results/fit-order246} give the main
results of my computations for the self-force and its error estimates.

These results are fully consistent at the $0.1$~ppm level with the
highly accurate frequency-domain results of Detweiler, Messaritaki,
and Whiting~\cite{Detweiler-Messaritaki-Whiting-2003}.  Below
$0.1$~ppm my results' error estimates become increasingly unreliable
due to floating-point roundoff errors, and below $0.045$~ppm the 
finite accuracy of the Detweiler, Messaritaki, and
Whiting~\cite{Detweiler-Messaritaki-Whiting-2003} results
(their quoted error estimate is $0.015$~ppm) begins to affect
comparisons with my results.

\begingroup
\squeezetable
\begin{table*}[bp]
\begin{flushleft}
Tail fit fits coefficients $\{c_4,c_6\}$ ($c_2$ is given analytically
by the circular-orbit formula~\eqref{eqn-c_2-tail-fit-coeff}):		\\[1ex]
\begin{ruledtabular}
\begin{tabular}{lllldldldld}
Floating-point	&
		&		&
				& \multicolumn{1}{c}{$\chi^2$}
				& \multicolumn{2}{c}{$\delta F_\self^\internal$}
				& \multicolumn{2}{c}{$\delta F_\self^\rp$}
				& \multicolumn{2}{c}{$\delta F_\self^\actual$}
									\\
\cline{6-7} \cline{8-9} \cline{10-11}					\\[-2ex]
Precision	&
$\AMRtolerance$	& AMR/FMR	& \multicolumn{1}{c}{$F_\self$}
				& \multicolumn{1}{c}{(11~d.o.f.)}
				& \multicolumn{1}{c}{absolute}
						& \multicolumn{1}{c}{relative}
				& \multicolumn{1}{c}{absolute}
						& \multicolumn{1}{c}{relative}
				& \multicolumn{1}{c}{absolute}
						& \multicolumn{1}{c}{relative}
									\\
\hline									\\[-2ex]
double		&
$10^{-14}$      & record        & $\SciNum{1.380\,7\Z\Z\,\Z\Z\Z\,\Z}{-5}$
				& 0.0889 
				& $\SciNum{2.2}{-8}$	& 1600
				& $\SciNum{2.2}{-8}$	& 1600
				& $\SciNum{2.2}{-8}$	& 1600
									\\
double		&
$10^{-14}$	& playback2	& $\SciNum{1.378\,59\Z\,\Z\Z\Z\,\Z}{-5}$
				& 0.0912
				& $\SciNum{1.4}{-9}$	& 100
				&			&
				& $\SciNum{1.4}{-9}$	& 100
									\\[1ex]
double		&
$10^{-15}$	& record	& $\SciNum{1.378\,73\Z\,\Z\Z\Z\,\Z}{-5}$
				& 0.109
				& $\SciNum{2.2}{-9}$	& 160
				& $\SciNum{2.8}{-9}$	& 210
				& $\SciNum{2.8}{-9}$	& 210
									\\
double		&
$10^{-15}$	& playback2	& $\SciNum{1.378\,466\,\Z\Z\Z\,\Z}{-5}$
				& 0.113
				& $\SciNum{1.4}{-10}$	& 10
				&			&
				& $\SciNum{1.8}{-10}$	& 13
									\\
double		&
$10^{-15}$	& playback3	&
				&
				&			&
				&			&
				&			&
									\\
double		&
$10^{-15}$	& playback4	&
				&
				&			&
				&			&
				&			&
									\\[1ex]
double		&
$10^{-16}$	& record	& $\SciNum{1.378\,541\,\Z\Z\Z\,\Z}{-5}$
				& 0.192
				& $\SciNum{9.5}{-10}$	& 69
				& $\SciNum{9.2}{-10}$	& 67
				& $\SciNum{9.2}{-10}$	& 67
									\\
double		&
$10^{-16}$	& playback2	& $\SciNum{1.378\,453\,9\Z\Z\,\Z}{-5}$
				& 0.208 
				& $\SciNum{5.9}{-11}$	& 4.3
				& $\SciNum{5.5}{-11}$	& 4.0
				& $\SciNum{5.6}{-11}$	& 4.1
									\\
double		&
$10^{-16}$	& playback3	& $\SciNum{1.378\,449\,5\Z\Z\,\Z}{-5}$
				& 0.114
				& $\SciNum{1.2}{-11}$	& 0.85
				& $\SciNum{1.1}{-11}$	& 0.81
				& $\SciNum{1.2}{-11}$	& 0.90
									\\
double		&
$10^{-16}$	& playback4	& $\SciNum{1.378\,448\,76\Z\,\Z}{-5}$
				& 0.911
				& $\SciNum{3.9}{-12}$	& 0.29
				& $\SciNum{6.6}{-12}$	& 0.48
				& $\SciNum{4.8}{-12}$	& 0.35
									\\
double		&
$10^{-16}$	& playback6	& $\SciNum{1.378\,448\,23\Z\,\Z}{-5}$
				& 4.28
				& $\SciNum{1.6}{-12}$	& 0.11
				& $\SciNum{2.3}{-12}$	& 0.17
				& $\SciNum{5.2}{-13}$	& 0.038
									\\
\shaderow
double		&
$10^{-16}$	& playback8	& $\SciNum{1.378\,448\,39\Z\,\Z}{-5}$
				& 27.8
				& $\SciNum{1.6}{-12}$	& 0.12
				& 			&
				& $\SciNum{1.1}{-12}$	& 0.078
									\\
\hline									\\[-2ex]
long-double	&
$10^{-14}$      & record        & $\SciNum{1.380\,7\Z\Z\,\Z\Z\Z\,\Z}{-5}$
				& 0.0893 
				& $\SciNum{2.2}{-8}$	& 1600
				& $\SciNum{2.2}{-8}$	& 1600
				& $\SciNum{2.2}{-8}$	& 1600
									\\
long-double	&
$10^{-14}$	& playback2	& $\SciNum{1.378\,59\Z\,\Z\Z\Z\,\Z}{-5}$
				& 0.0915
				& $\SciNum{1.4}{-9}$	& 100
				&			&
				& $\SciNum{1.4}{-9}$	& 100
									\\[1ex]
long-double	&
$10^{-15}$	& record	& $\SciNum{1.378\,73\Z\,\Z\Z\Z\,\Z}{-5}$
				& 0.109
				& $\SciNum{2.2}{-9}$	& 160
				& $\SciNum{2.8}{-9}$	& 210
				& $\SciNum{2.8}{-9}$	& 210
									\\
long-double	&
$10^{-15}$	& playback2	& $\SciNum{1.378\,466\,\Z\Z\Z\,\Z}{-5}$
				& 0.113
				& $\SciNum{1.4}{-10}$	& 10
				&			&
				& $\SciNum{1.8}{-10}$	& 13
									\\[1ex]
long-double	&
$10^{-16}$	& record	& $\SciNum{1.378\,540\,\Z\Z\Z\,\Z}{-5}$
				& 0.193
				& $\SciNum{9.5}{-10}$	& 69
				& $\SciNum{9.2}{-10}$	& 67
				& $\SciNum{9.2}{-10}$	& 67
									\\
long-double	&
$10^{-16}$	& playback2	& $\SciNum{1.378\,454\,0\Z\Z\,\Z}{-5}$
				& 0.190
				& $\SciNum{6.0}{-11}$	& 4.3
				&			&
				& $\SciNum{5.7}{-11}$	& 4.1
									\\[1ex]
long-double	&
$10^{-17}$	& record	& $\SciNum{1.378\,461\,\Z\Z\Z\,\Z}{-5}$
				& 0.119 
				& $\SciNum{1.0}{-10}$	& 7.6
				& $\SciNum{1.3}{-10}$	& 9.6
				& $\SciNum{1.3}{-10}$	& 9.5
									\\
long-double	&
$10^{-17}$	& playback2	& $\SciNum{1.378\,448\,96\Z\,\Z}{-5}$
				& 0.180
				& $\SciNum{6.5}{-12}$	& 0.47
				&			&
				& $\SciNum{6.8}{-12}$	& 0.49
									\\[1ex]
long-double	&
$10^{-18}$	& record	& $\SciNum{1.378\,451\,3\Z\Z\,\Z}{-5}$
				& 0.200 
				& $\SciNum{3.1}{-11}$	& 2.3
				& $\SciNum{3.1}{-11}$	& 2.2
				& $\SciNum{3.0}{-11}$	& 2.2
									\\
long-double	&
$10^{-18}$	& playback2	& $\SciNum{1.378\,448\,377\,\Z}{-5}$
				& 0.238 
				& $\SciNum{2.0}{-12}$	& 0.14
				&			&
				& $\SciNum{9.7}{-13}$	& 0.070
									\\
long-double	&
$10^{-19}$	& record	& $\SciNum{1.378\,448\,82\Z\,\Z}{-5}$
				& 0.178
				& $\SciNum{6.0}{-12}$	& 0.44
				& $\SciNum{7.0}{-12}$	& 0.51
				& $\SciNum{5.4}{-12}$	& 0.39
									\\
long-double	&
$10^{-19}$	& playback2	& $\SciNum{1.378\,448\,169\,8}{-5}$
				& 5.06
				& $\SciNum{3.7}{-13}$	& 0.027
				& $\SciNum{9.8}{-14}$	& 0.0071
				& $\SciNum{1.1}{-12}$	& 0.080
									\\
\shaderow
long-double	&
$10^{-19}$	& playback23	& $\SciNum{1.378\,448\,177\,7}{-5}$
				& 69.1
				& $\SciNum{9.8}{-14}$	& 0.0071
				&			&
				& $\SciNum{1.0}{-12}$	& 0.074
									\\
\hline									\\[-2ex]
\multicolumn{3}{l}{Detweiler, Messaritaki, and Whiting}
				& $\SciNum{1.378\,448\,28\Z\,\Z}{-5}$
				&
				& $\SciNum{2\P{.0}}{-13}$ & 0.015
				&			&
				& $\CenterInSciNumSpace{(0)}{0.0}{-00}$
			& \multicolumn{1}{l}{\Z\Z\Z\CenterWithSizeOf{$(0)$}{0}}
									\\
\end{tabular}
\end{ruledtabular}
\end{flushleft}
\caption[Self-Force Results and Error Estimates, Fitting $\{c_4,c_6\}$]
	{
	This table shows the main results of the self-force calculations
	for the case where the tail fit fits only $\{c_4,c_6\}$.
	For each calculation, the table shows the AMR error
	tolerance~$\AMRtolerance$ used in numerically solving
	the wave equation~\eqref{eqn-wave},
	whether this numerical solution is purely AMR (``record'')
	or also uses FMR (``playback$N$'' for some $N$), the
	computed self-force $F_\self$, $\chi^2$ for the tail fit,
	the internal error estimate~$\delta F_\self^\internal$,
	the record-playback error estimate~$\delta F_\self^\rp$,
	and the actual error~$\delta F_\self^\actual$.
	Each error estimate or error is shown both as an absolute
	value, and as a relative value in parts per million (ppm)
	relative to the overall self-force.
	The shaded rows have very large tail-fit~$\chi^2$, so their
	internal estimates may be unreliable.
	For comparison, the last row of this table shows the highly accurate
	frequency-domain value calculated by Detweiler, Messaritaki,
	and Whiting~\cite{Detweiler-Messaritaki-Whiting-2003}.
	}
\label{tab-self-force-and-error-results/fit-order46}
\end{table*}
\endgroup

\begingroup
\squeezetable
\begin{table*}[bp]
\begin{flushleft}
Tail fit fits coefficients $\{c_2,c_4,c_6\}$:				\\[1ex]
\begin{ruledtabular}
\begin{tabular}{lllldldldld}
Floating-point	&
		&		&
				& \multicolumn{1}{c}{$\chi^2$}
				& \multicolumn{2}{c}{$\delta F_\self^\internal$}
				& \multicolumn{2}{c}{$\delta F_\self^\rp$}
				& \multicolumn{2}{c}{$\delta F_\self^\actual$}
									\\
\cline{6-7} \cline{8-9} \cline{10-11}					\\[-2ex]
Precision	&
$\AMRtolerance$	& AMR/FMR	& \multicolumn{1}{c}{$F_\self$}
				& \multicolumn{1}{c}{(10~d.o.f.)}
				& \multicolumn{1}{c}{absolute}
						& \multicolumn{1}{c}{relative}
				& \multicolumn{1}{c}{absolute}
						& \multicolumn{1}{c}{relative}
				& \multicolumn{1}{c}{absolute}
						& \multicolumn{1}{c}{relative}
									\\
\hline									\\[-2ex]
double		&
$10^{-14}$      & record        & $\SciNum{1.381\,0\Z\Z\,\Z\Z\Z}{-5}$
				& 0.0877
				& $\SciNum{1.0}{-7}$	& 7300
				& $\SciNum{2.6}{-8}$	& 1900
				& $\SciNum{2.6}{-8}$	& 1900
									\\
double		&
$10^{-14}$	& playback2	& $\SciNum{1.378\,61\Z\,\Z\Z\Z}{-5}$
				& 0.0901
				& $\SciNum{6.3}{-9}$	& 460
				&			&
				& $\SciNum{1.6}{-9}$	& 120
									\\[1ex]
double		&
$10^{-15}$	& record	& $\SciNum{1.379\,39\Z\,\Z\Z\Z}{-5}$
				& 0.0225
				& $\SciNum{2.2}{-8}$	& 1600
				& $\SciNum{9.4}{-9}$	& 680
				& $\SciNum{9.4}{-9}$	& 680
									\\
double		&
$10^{-15}$	& playback2	& $\SciNum{1.378\,508\,\Z\Z\Z}{-5}$
				& 0.0233
				& $\SciNum{1.4}{-9}$	& 100
				&			&
				& $\SciNum{6.0}{-10}$	& 43
									\\
double		&
$10^{-15}$	& playback3	&
				&
				&			&
				&			&
				&			&
									\\
double		&
$10^{-15}$	& playback4	&
				&
				&			&
				&			&
				&			&
									\\[1ex]
double		&
$10^{-16}$	& record	& $\SciNum{1.378\,473\,\Z\Z\Z}{-5}$
				& 0.177
				& $\SciNum{5.5}{-9}$	& 400
				& $\SciNum{2.4}{-10}$	& 18
				& $\SciNum{2.4}{-10}$	& 18
									\\
double		&
$10^{-16}$	& playback2	& $\SciNum{1.378\,449\,8\Z\Z}{-5}$
				& 0.193
				& $\SciNum{3.5}{-10}$	& 25
				& $\SciNum{5.2}{-12}$	& 0.38
				& $\SciNum{1.6}{-11}$	& 1.1
									\\
double		&
$10^{-16}$	& playback3	& $\SciNum{1.378\,449\,43\Z}{-5}$
				& 0.114
				& $\SciNum{6.8}{-11}$	& 4.9
				& $\SciNum{7.2}{-12}$	& 0.52
				& $\SciNum{1.2}{-11}$	& 0.84
									\\
double		&
$10^{-16}$	& playback4	& $\SciNum{1.378\,448\,94\Z}{-5}$
				& 0.904
				& $\SciNum{2.2}{-11}$	& 1.6
				& $\SciNum{5.0}{-12}$	& 0.36
				& $\SciNum{6.6}{-12}$	& 0.48
									\\
double		&
$10^{-16}$	& playback6	& $\SciNum{1.378\,448\,54\Z}{-5}$
				& 3.27
				& $\SciNum{3.5}{-12}$	& 0.25
				& $\SciNum{2.4}{-11}$	& 1.7
				& $\SciNum{2.6}{-12}$	& 0.19
									\\
\shaderow
double		&
$10^{-16}$	& playback8	& $\SciNum{1.378\,446\,91\Z}{-5}$
				& 24.5
				& $\SciNum{8.3}{-12}$	& 0.60
				&			&
				& $\SciNum{1.4}{-11}$	& 1.0
									\\
\hline									\\[-2ex]
long-double	&
$10^{-14}$      & record        & $\SciNum{1.381\,0\Z\Z\,\Z\Z\Z}{-5}$
				& 0.0880
				& $\SciNum{1.0}{-7}$	& 7300
				& $\SciNum{2.6}{-8}$	& 1900
				& $\SciNum{2.6}{-8}$	& 1900
									\\
long-double	&
$10^{-14}$	& playback2	& $\SciNum{1.378\,61\Z\,\Z\Z\Z}{-5}$
				& 0.0903
				& $\SciNum{6.3}{-9}$	& 460
				&			&
				& $\SciNum{1.6}{-9}$	& 120
									\\[1ex]
long-double	&
$10^{-15}$	& record	& $\SciNum{1.379\,39\Z\,\Z\Z\Z}{-5}$
				& 0.0225
				& $\SciNum{2.2}{-8}$	& 1600
				& $\SciNum{9.4}{-9}$	& 680
				& $\SciNum{9.4}{-9}$	& 680
									\\
long-double	&
$10^{-15}$	& playback2	& $\SciNum{1.378\,509\,\Z\Z\Z}{-5}$
				& 0.0227
				& $\SciNum{1.4}{-9}$	& 100
				&			&
				& $\SciNum{6.0}{-10}$	& 44
									\\[1ex]
long-double	&
$10^{-16}$	& record	& $\SciNum{1.378\,473\,\Z\Z\Z}{-5}$
				& 0.177
				& $\SciNum{5.5}{-9}$	& 400
				& $\SciNum{2.4}{-10}$	& 17
				& $\SciNum{2.4}{-10}$	& 18
									\\
long-double	&
$10^{-16}$	& playback2	& $\SciNum{1.378\,450\,1\Z\Z}{-5}$
				& 0.178
				& $\SciNum{3.5}{-10}$	& 25
				&			&
				& $\SciNum{1.9}{-11}$	& 1.3
									\\[1ex]
long-double	&
$10^{-17}$	& record	& $\SciNum{1.378\,505\,\Z\Z\Z}{-5}$
				& 0.0238
				& $\SciNum{1.4}{-9}$	& 100
				& $\SciNum{5.6}{-10}$	& 41
				& $\SciNum{5.7}{-10}$	& 41
									\\
long-double	&
$10^{-17}$	& playback2	& $\SciNum{1.378\,452\,5\Z\Z}{-5}$
				& 0.0191
				& $\SciNum{8.9}{-11}$	& 6.4
				&			&
				& $\SciNum{4.2}{-11}$	& 3.1
									\\[1ex]
long-double	&
$10^{-18}$	& record	& $\SciNum{1.378\,453\,2\Z\Z}{-5}$
				& 0.196
				& $\SciNum{2.9}{-10}$	& 21
				& $\SciNum{4.6}{-11}$	& 3.3
				& $\SciNum{4.9}{-11}$	& 3.6
									\\
long-double	&
$10^{-18}$	& playback2	& $\SciNum{1.378\,448\,92\Z}{-5}$
				& 0.148
				& $\SciNum{1.8}{-11}$	& 1.3
				&			&
				& $\SciNum{6.4}{-12}$	& 0.47
									\\
long-double	&
$10^{-19}$	& record	& $\SciNum{1.378\,450\,1\Z\Z}{-5}$
				& 0.0907
				& $\SciNum{4.4}{-11}$	& 3.2
				& $\SciNum{1.5}{-11}$	& 1.1
				& $\SciNum{1.8}{-11}$	& 1.3
									\\
long-double	&
$10^{-19}$	& playback2	& $\SciNum{1.378\,448\,69\Z}{-5}$
				& 1.23
				& $\SciNum{2.7}{-12}$	& 0.19
				& $\SciNum{1.8}{-12}$	& 0.13
				& $\SciNum{4.1}{-12}$	& 0.29
									\\
long-double	&
$10^{-19}$	& playback23	& $\SciNum{1.378\,448\,541}{-5}$
				& 12.0
				& $\SciNum{4.9}{-13}$	& 0.036
				&			&
				& $\SciNum{2.6}{-12}$	& 0.19
									\\
\hline									\\[-2ex]
\multicolumn{3}{l}{Detweiler, Messaritaki, and Whiting}
				& $\SciNum{1.378\,448\,28\Z}{-5}$
				&
				& $\SciNum{2\P{.0}}{-13}$ & 0.015
				&			&
				& $\CenterInSciNumSpace{(0)}{0.0}{-00}$
			& \multicolumn{1}{l}{\Z\Z\Z\CenterWithSizeOf{$(0)$}{0}}
									\\
\end{tabular}
\end{ruledtabular}
\end{flushleft}
\caption[Self-Force Results and Error Estimates, Fitting $\{c_2,c_4,c_6\}$]
	{
	This table shows the main results of the self-force calculations
	for the case where the tail fit fits $\{c_2,c_4,c_6\}$, as might
	be the case for a non-circular particle orbit.
	For each calculation, the table shows the AMR error
	tolerance~$\AMRtolerance$ used in numerically solving
	the wave equation~\eqref{eqn-wave},
	whether this numerical solution is purely AMR (``record'')
	or also uses FMR (``playback$N$'' for some $N$), the
	computed self-force $F_\self$, $\chi^2$ for the tail fit,
	the internal error estimate~$\delta F_\self^\internal$,
	the record-playback error estimate~$\delta F_\self^\rp$,
	and the actual error~$\delta F_\self^\actual$.
	Each error estimate or error is shown both as an absolute
	value, and as a relative value in parts per million (ppm)
	relative to the overall self-force.
	The shaded rows have very large tail-fit~$\chi^2$, so their
	internal estimates may be unreliable.
	For comparison, the last row of this table shows the highly accurate
	frequency-domain value calculated by Detweiler, Messaritaki,
	and Whiting~\cite{Detweiler-Messaritaki-Whiting-2003}.
	}
\label{tab-self-force-and-error-results/fit-order246}
\end{table*}
\endgroup


\section{Conclusions}
\label{sect-conclusions}

This work demonstrates that the use of characteristic AMR can
dramatically improve the efficiency of time-domain self-force
calculations using the Barack-Ori mode-sum regularization formalism.
\footnote{
	 Ca\~{n}izares and Sopuerta~\cite{Canizares-Sopuerta-2009a,
Canizares-Sopuerta-2009b} have used a multi-domain pseudospectral
	 method to numerically solve the wave equation~\eqref{eqn-wave}
	 within the same Barack-Ori mode-sum regularization
	 framework as used here.  Their results are quite
	 promising, but are of relatively low accuracies
	 (with typical relative errors of $\sim 10^{-3}$)
	 compared to the results reported here.
	 }

I find that the tail-fit basis $\{f_p\}$ is very ill-conditioned if
many terms in the tail series~\eqref{eqn-tail-series} are fit simultaneously.
Fortunately, normalizing the basis functions to have similar magnitudes
mostly alleviates this ill-conditioning.

Past self-force calculations have often used ``record-playback''
error estimates derived from comparing a pair of different-resolution
calculations.  Here I present, and validate as quite reliable, a
set of internal error estimators which can be used within a single
self-force calculation (whether AMR or unigrid) to estimate the
accuracy of individual regularized self-force modes $F_{\ell,\reg}$,
and the numerical force, tail force, and overall self-force derived
from them.

In their pioneering calculation of the gravitational self-force
acting on a mass particle on a circular geodesic orbit in Schwarzschild
spacetime, Barack and Sago~\cite{Barack-Sago-2010} use the arithmetic-sum
formula~\eqref{eqn-numerical-force-internal-error-estimate/arithmetic-sum}
to combine the numerical-force error given the (record-playback)
error estimates of the individual modes.  Here I show that (at least
for my results) this formula is unnecessarily conservative, and
that the quadrature-sum
formula~\eqref{eqn-numerical-force-internal-error-estimate/quadrature-sum}
provides a better approximation to the numerical-force error over
a wide range of overall computational accuracies.

Like other researchers (see, for example,
\cite{Sago-Barack-Detweiler-2008,Barack-Sago-2010}
and references therein, and the references cited in
footnote~\ref{footnote-cmp-different-calculations}),
I find excellent agreement between time- and frequency-domain
calculations of the self-force.   Here I demonstrate this agreement
down to the $0.1$~ppm accuracy level.  Because the time- and
frequency-domain calculations are structured so differently, this
high-precision verification of their agreement provides a strong
confirmation of the correctness of both calculations, and implicitly
of their respective theoretical formalisms as well.

The present work could (should) be extended in several directions.
Apart from the technical limits of the relatively coarse adjustment
of the problem-domain size, one obvious extension would be to consider
the electromagnetic and/or gravitational self-force.  Another
possibility would be to generalize the current finite differencing
scheme to handle non-circular particle orbits.  This would be
straightforward, albeit somewhat tedious, using techniques such as
those described by Haas~\cite{Haas-2007} and
Barack and Sago~\cite{Barack-Sago-2010}.  Once non-circular orbits
can be handled, it should then be possible to move to full
orbit-correction calculations of the type suggested by
Gralla and Wald~\cite[section~7]{Gralla-Wald-2008}.
This will be very computationally demanding (it might benefit from
further increasing the order of the finite differencing), but
should provide valuable information about $\O(\mu^2)$~radiation-reaction
effects.

The generalization of this work to particle orbits in Kerr spacetime
would also be very valuable, but would demand a major reorganization
of the mathematical and computational structure.


\begin{acknowledgments}
I think Leor Barack for introducing me to the self-force problem,
and Leor Barack, Norichika Sago, and Darren Golbourn for many
valuable conversations and assistance with the self-force calculations.
I thank Eric Ost for valuable assistance with the computer cluster
used for many of the calculations described in this paper.
I thank Michael Trosset, Stephanie Dickinson, and Lijiang Guo
of Indiana University's Indiana Statistical Consulting Center for
their advice concerning the ill-conditioning and scaling of the tail fit.
\end{acknowledgments}

\end{document}